\documentclass[lineno]{jfm}

\usepackage{graphicx}
\usepackage{multirow}
\usepackage{newtxtext}
\usepackage{newtxmath}
\usepackage{natbib}
\usepackage{hyperref}
\hypersetup{
    colorlinks = true,
    urlcolor   = blue,
    citecolor  = black,
}
\newcommand{\RomanNumeralCaps}[1]
\linenumbers
\usepackage{longtable}

\title{Asymmetric velocity boundary conditions lead to zonal flow in centrifugal convection}

\author{Jun Zhong\aff{1}
 \and Chao Sun\aff{1,2}
 \corresp{\email{chaosun@tsinghua.edu.cn}}}

\affiliation{\aff{1} New Cornerstone Science Laboratory, Center for Combustion Energy, Key Laboratory for
Thermal Science and Power Engineering of Ministry of Education, Department of Energy and
Power Engineering, Tsinghua University, 100084 Beijing, China
\aff{2} Department of Engineering Mechanics, School of Aerospace Engineering, Tsinghua University, 100084 Beijing, China}

\begin{document}
\maketitle

\begin{abstract}
We perform direct numerical simulations of rapidly rotating annular centrifugal convection to investigate how mixed (asymmetric) velocity boundary conditions and geometric curvature shape the flow organisation and heat transfer. Motivated by the quasi-two-dimensionalisation under strong rotation and the long spin-up required for large-scale states, we employ two-dimensional simulations and consider four boundary-condition sets: no-slip/no-slip (INON), no-slip/stress-free (INOS), stress-free/no-slip (ISON) and stress-free/stress-free (ISOS). For fixed geometry with the radius ratio $\eta=0.5$ and over the Rayleigh number $Ra\in[10^6,10^9]$, the heat transfer is strongest for ISOS, followed by INOS and INON, while ISON exhibits a pronounced suppression as a strong zonal flow aligned with the rotation develops. In the three cases dominated by large-scale circulation, the Nusselt number $Nu$ follows an effective classical-type scaling close to $Nu\sim Ra^{0.27}$, whereas the zonal-flow branch displays a much weaker scaling $Nu\sim Ra^{0.1}$ and strong flow anisotropy with the large difference between the radial and azimuthal Reynolds numbers $Re_r\ll Re_\varphi$. A dissipation analysis shows that zonal-flow formation is accompanied by a transition from boundary-layer-dominated dissipation to a relatively low and uniform bulk dissipation, consistent with shear-induced plume suppression. By varying the radius ratio $\eta$, we demonstrate that increasing $\eta$ weakens curvature asymmetry and destabilises the zonal-flow state, leading to roll-dominated convection in the planar limit, and we relate the accompanying bulk-temperature asymmetry to the boundary heat flux asymmetry using a free-convective boundary-layer model.
 
\end{abstract}


\section{Introduction}

Thermal convection is a ubiquitous phenomenon in natural environments, playing an important role in many geophysical and astrophysical systems \citep{1929Tropical, hanasoge2016seismic, wicht2019advances, hartmann_toward_2024}. Fluids subjected to uneven heating under gravitational fields generate motion due to buoyancy, with heat being converted into kinetic energy driving large-scale turbulent motion. As one typical paradigm of thermal convection, Rayleigh-B\'enard convection (RBC), wherein the flow is heated below and cooled above, has attracted considerable attention over the past few decades \citep{grossmann_scaling_2000, ahlers_heat_2009, lohse_small-scale_2010, chilla_new_2012, yeh_geophysical_2023, lohse_rmp_2024}. In the traditional planar RBC system, regular large-scale circulation (LSC) formed by upward-moving hot plumes and downward-moving cold plumes, is the primary flow structure, continuously transferring heat from the lower plate to the upper plate \citep{grossmann_scaling_2000, grossmann_thermal_2001, xia_current_2013}. In other variants of RBC, such as Centrifugal Convection (CC) in the annular structure \citep{jiang_supergravitational_2020, rouhi_coriolis_2021, jiang_experimental_2022, bhadra_boundary-layer_2024, bhadra_heat_2025, yao2025direct, lai_gravity_2025} and turbulent spherical RBC in the spherical shell \citep{fan_scaling_2024, fu_turbulent_2024}, the LSC morphology remains well-preserved and continues to serve as the primary structure for heat transfer within the classical regime.

In real-world natural and industrial settings, thermal convection typically develops under complex conditions, in which multiple factors can disrupt the large-scale circulation and thereby alter heat transfer efficiency. Externally imposed flows—such as Couette flows driven by boundary motion or pressure-driven Poiseuille flows—can effectively disrupt LSC patterns \citep{blass_flow_2020, leng_flow_2021, zhong_sheared_2023, yerragolam_scaling_2024, zhong_centrifugal_2025}. Under such background shear, vertical plumes become deflected, large-scale vortices are stretched and eventually break up, plume generation is hindered, and the overall heat transfer efficiency is notably reduced \citep{yerragolam_how_2022,zhong_effect_2024}. Rotational effects can also modify the flow structure through the action of the Coriolis force. According to the Taylor–Proudman constraint, when rotation is sufficiently strong, the primary flow in the Rotating Rayleigh–Bénard convection develops into vertically aligned Taylor columns that connect the top and bottom plates \citep{king_boundary_2009, king_turbulent_2013, kunnen_role_2011, kunnen_structure_2013, ecke_2023_turbulent, song_direct_2024, song_scaling_2024}. The presence of a porous medium may suppress thermal convection as well. When the medium is relatively sparse, large-scale flow structures remain largely intact; however, as its density increases and permeability decreases, the accompanying rise in viscous resistance destabilizes these structures, leading heat transfer to occur predominantly through individual plume flows \citep{liu2020rayleigh, xu_pore-scale_2023, zhong_2023, zhu_transport_2024, li_darcy_2025}.

In addition, another crucial factor that governs flow patterns and heat transfer is the boundary condition of the system. In the canonical Rayleigh–B\'enard convection setup, no-slip and isothermal boundary conditions are imposed on both the top and bottom plates, while periodic conditions are applied along the horizontal directions. Temporal or spatial modulations of the boundary temperature can influence plume fluctuations and spacing, and even significantly enhance heat transfer efficiency through carefully designed temperature distributions \citep{yang_periodically_2020, zhao_modulation_2022, zhou_deep_2025}. Regarding velocity boundary conditions, stress-free boundaries provide a favorable environment for the emergence of zonal flows owing to their reduced viscous dissipation. Two distinct flow states—zonal flow and convection rolls—have been identified in two-dimensional Rayleigh–B\'enard convection with stress-free plates and periodic lateral boundaries. These two states were initially found to coexist under identical parameter conditions and were shown to be strongly influenced by the initial state and the system’s aspect ratio \citep{goluskin_convectively_2014, van_der_poel_effect_2014, wen_steady_2020, wang_zonal_2020}. Through extensive direct numerical simulations with prolonged integration times, \cite{wang_lifetimes_2023} further demonstrated that zonal flow is metastable at moderate Rayleigh numbers, eventually decaying into the roll state after a finite lifetime with survival probability decreasing exponentially in time.

However, in geophysical and astrophysical systems, zonal flows are commonly observed to persist over long times, for example in the global atmospheric circulation and in Jupiter’s zonal winds \citep{Schneider2006The, yadav_deep_2020, heimpel_polar_2022}. This contrast raises the question of what additional mechanisms control the formation and long-term stability of zonal flow. In particular, it remains unclear how rotation and geometric curvature, and their interplay, shape the emergence of zonal flow and the associated heat transfer. Moreover, real-world flows often involve mixed velocity boundary conditions, with one boundary no-slip and the other stress-free, as may be relevant to near-surface atmospheric motions and oceanic currents. These considerations motivate a more systematic investigation of zonal-flow dynamics and heat transport under mixed boundary conditions.

Therefore, this paper focuses on thermal convection subject to mixed (asymmetric) velocity boundary conditions, and examines how rotation and curvature jointly influence the resulting flow and heat transfer. For simplicity, we consider an annular CC configuration, in which the flow under strong rotation is constrained to be quasi-two-dimensional. We investigate four sets of velocity boundary conditions in the presence of rotation and curvature: no-slip at both boundaries, stress-free at both boundaries, no-slip at the inner boundary with stress-free at the outer boundary, and stress-free at the inner boundary with no-slip at the outer boundary. In addition, the role of curvature is assessed by varying the radius ratio of the annulus.

The rest of the paper is organized as follows: the establishment of the numerical model and the direct numerical simulations are introduced in section \ref{sec2}, and the main results are thoroughly discussed in section \ref{sec3}. Finally, conclusions are presented in section \ref{sec4}.

\section{Numerical Model} \label{sec2}

\begin{figure}
    \centering
    \includegraphics[width=1.0\linewidth]{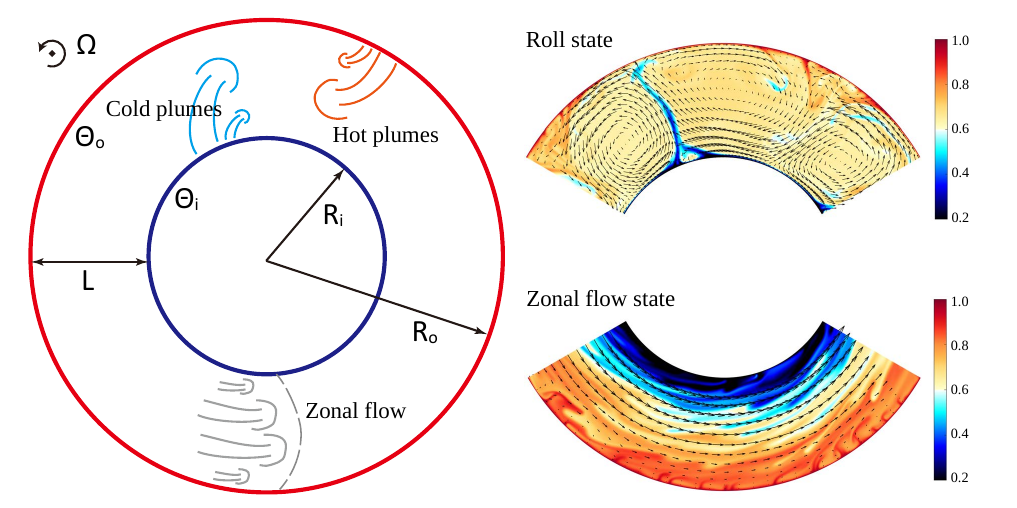}
    \captionsetup{justification=justified}
    \caption{ Left: schematic diagram of the flow configuration in the stationary reference frame with angular velocity $\Omega$. $R_{i,o}$ and $\Theta_{i,o}$ are the radius and temperature of the inner and outer cylinders, respectively. Hot and cold plumes are generated in the corresponding boundaries and form the roll state, while the strong azimuthal zonal flow leads to the zonal flow state. Right: two typical instantaneous temperature snapshots with velocity vectors are displayed to demonstrate the different flow characteristics between the roll state (upper figure) and the zonal flow state (lower figure).}
	\label{fig: Schema}
\end{figure}

As the flow in the CC system becomes quasi-two-dimensional under strong rotation \citep{jiang_supergravitational_2020, zhong_sheared_2023}, and since resolving the evolution of zonal flow may require prolonged integration times \citep{wang_lifetimes_2023}, we adopt two-dimensional simulations to reduce the computational cost. This choice is further supported by evidence that, under rotation-constrained conditions, the three-dimensional zonal-flow dynamics in RBC remain consistent with their two-dimensional counterpart \citep{wang_zonal_2020}, making two-dimensional simulations a reliable alternative for the corresponding three-dimensional computations. A two-dimensional CC system rotating with angular velocity $\Omega$ in a cylindrical coordinate $(r,\varphi)$ is illustrated in figure \ref{fig: Schema}. The incompressible viscous fluid is bounded by a cold inner cylinder with radius $R_i$ and a hot outer cylinder with radius $R_o$. The gap between the two cylinders is $L=R_o-R_i$. $\Theta_{i,o}$ is the temperature of inner and outer cylinders, respectively, and the temperature difference is $\Delta=\Theta_o-\Theta_i$. Some fluid properties, including the coefficient of thermal expansion $\alpha$, the kinematic viscosity $\nu$, and the thermal diffusivity $\kappa$, are assumed to be constant in the system. 

\subsection{Governing equations}

In the CC configuration, the fluid is heated at the outer wall and, driven by centrifugal buoyancy, moves radially inward toward the inner wall. The free-fall velocity is defined based on the mean centrifugal acceleration as $U=\sqrt{\alpha\Delta L\Omega^2(R_i+R_o)/2}$. Using the gap width $L$, the free-fall velocity $U$, and the imposed temperature difference $\Delta$ as the characteristic length, velocity, and temperature scales, respectively, the non-dimensional governing equations under the Oberbeck–Boussinesq assumption can be written as \citep{jiang_supergravitational_2020, jiang_experimental_2022, zhong_sheared_2023, zhong_effect_2024}:

\begin{equation}\label{eq: OB}
    \begin{aligned}
    \boldsymbol{\nabla}\cdot\boldsymbol{u}&=0,\\
    \frac{\partial\boldsymbol{u}}{\partial t}+\boldsymbol{u}\cdot\boldsymbol{\nabla u}=-\boldsymbol{\nabla} p-Ro^{-1}\boldsymbol{e_z}\times\boldsymbol{u}&+\sqrt{\frac{Pr}{Ra}}{\boldsymbol{\nabla}^2}\boldsymbol{u}-\theta\frac{2(1-\eta)}{1+\eta}\boldsymbol{r},\\
    \frac{\partial \theta}{\partial t}+\boldsymbol{\nabla}\cdot (\boldsymbol{u}\theta)&=\sqrt{\frac{1}{Ra\cdot Pr}}\boldsymbol{\nabla}^2 \theta,\\
    \end{aligned}
\end{equation}
where $\boldsymbol{u}=(u_r,u_\varphi,u_z)$ is the velocity vector, $p$ is the pressure (the density is contained within), $\theta$ is the temperature,  $\boldsymbol{e_z}$ is the unit vector in the axial direction and $\eta=R_i/R_o$ is the radius ratio. Three dimensionless parameters appearing in the governing equations are the Rayleigh number $Ra$, the inverse Rossby number $Ro^{-1}$ and the Prandtl number $Pr$, which read:
\begin{equation}\label{eq: Par}       
        Ra=\frac{\alpha\Delta{L}^3\Omega^{2}(R_i+R_o)/2}{\nu\kappa}, \quad Ro^{-1}=\frac{2\Omega L}{U},\quad
        Pr=\frac{\nu}{\kappa}.    
\end{equation}

In our system, isothermal boundary conditions are applied on the inner and outer boundaries. Due to the internal and external asymmetry of the cylindrical structure, there are four types of velocity boundary condition sets: no-slip inner and outer boundaries (INON), no-slip inner boundary and stress-free outer boundary (INOS), stress-free inner boundary and no-slip outer boundary (ISON), stress-free inner and outer boundaries (ISOS). By comparing the results from these four boundary conditions, it is hoped that the impact of asymmetry can be investigated in greater depth. 

\subsection{Direct Numerical Simulations}

To address this problem, we carry out a series of direct numerical simulations (DNS). The governing equations \eqref{eq: OB} are solved with the energy-conserving, second-order finite-difference solver AFiD on a staggered grid with Chebyshev-type clustering. Time integration is performed using a fractional-step third-order Runge–Kutta method, with the implicit contributions treated via a Crank–Nicolson scheme. AFiD has been extensively used and carefully validated in previous studies of RBC, Taylor–Couette flow, and plane–Couette flow \citep{verzicco_finite-difference_1996,van_der_poel_pencil_2015,zhu_afid-gpu_2018, wang_zonal_2020}.

All simulations are performed at sufficient spatial and temporal resolution, and a posteriori checks are conducted to confirm that all dynamically relevant scales are well resolved. Specifically, we evaluate the ratios of the maximum grid spacing $\Delta_g$ to the Kolmogorov length scale $\eta_K=(\nu^3/\varepsilon)^{1/4}$, where $\varepsilon$ is the mean viscous dissipation rate obtained via the exact relation, and to the Batchelor scale $\eta_B=\eta_KPr^{-1/2}$ \citep{silano_numerical_2010}; the results are reported in Appendix \ref{app2}. In addition, the clipped Chebyshev-type grid clustering in the radial direction provides adequate resolution of the boundary layers, ensuring enough grid points are placed within the thermal boundary layer. For the temporal resolution, we enforce the Courant–Friedrichs–Lewy (CFL) constraint and use $CFL\le 0.7$ to ensure numerical stability \citep{ostilla_optimal_2013,van_der_poel_pencil_2015,zhang_statistics_2017}. Random perturbations are applied as initial conditions, and considering the zonal flow state may be metastable and turn into a roll state \citep{wang_zonal_2020, wang_lifetimes_2023}, sufficient simulation time (larger than 5000 time units) is guaranteed for the flow in the zonal flow state to ensure stability. After the flow reaches a statistically stationary state, the simulations are continued for a sufficiently long time to achieve satisfactory statistical convergence. More numerical details are illustrated in Appendix \ref{app2}.

In the present study, we aim to investigate the effects of boundary condition sets and the curvature on the flow state and heat transfer. A wide range of $Ra$ over three decades, $Ra\in [10^6, 10^9]$, is focused on showing the effect on scalings. The other parameters are taken following previous experiments and numerical simulations on the CC system, with fixed $Ro^{-1}$ and $Pr=4.3$. In addition, to figure out the effect of curvature on the generation of the zonal flow, various radius ratios $\eta \in [0.3,1]$ are studied as well, where $\eta=1$ corresponds to the planar RBC system.

\section{Results and Discussion} \label{sec3}

\begin{figure}
    \centering
    \includegraphics[width=1.0\linewidth]{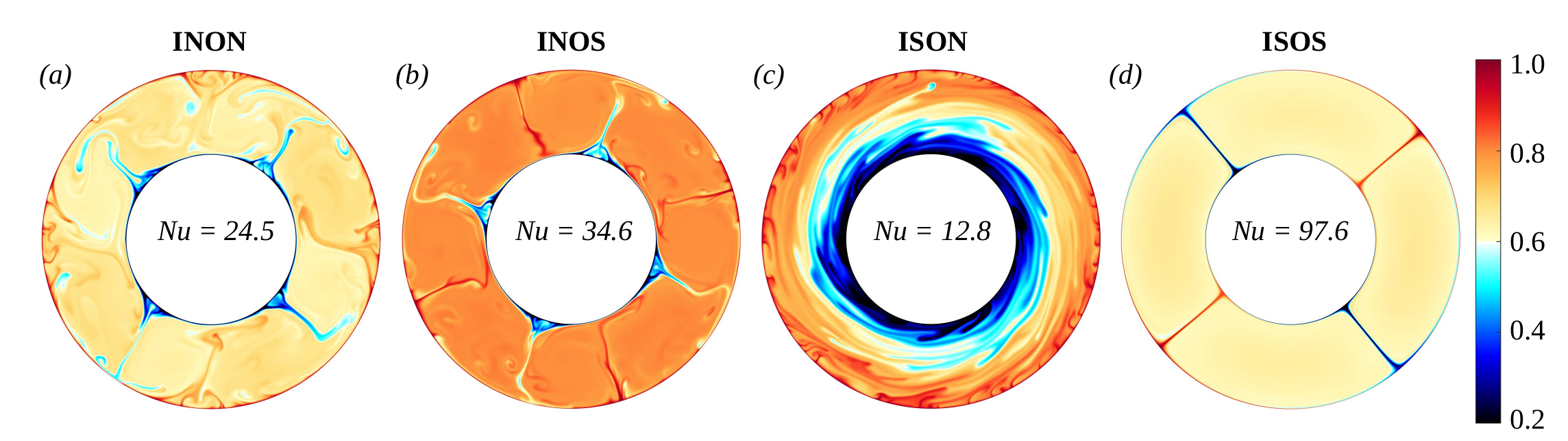}
    \captionsetup{justification=justified}
    \caption{ Typical instantaneous temperature snapshots and corresponding Nusselt number of different boundary condition sets: (a) INON, (b) INOS, (c) ISON, (d) ISOS. $Ra=10^8, \eta=0.5.$ }
	\label{fig: Tslices_type}
\end{figure}

Four typical snapshots of instantaneous temperature fields and corresponding time-averaged Nusselt numbers under different boundary condition sets with fixed geometry are shown in the figure \ref{fig: Tslices_type}. It is evident that both the flow structure and heat transfer efficiency undergo significant changes as boundary conditions are altered. For cases where both sides are no-slip (INON, fig. \ref{fig: Tslices_type}(a)), four pairs of convection rolls composed of one large and one small vortex are present in the system, which is influenced by the Coriolis force and the geometry limitation \citep{wang_effects_2022}. When the outer wall becomes a stress-free boundary (INOS, fig. \ref{fig: Tslices_type}(b)), the flow structure remains, while the bulk temperature increases a lot. The temperature differences between the bulk and the no-slip inner boundary wall lead to a $40\%$ increase in the Nusselt number as well. Under identical flow configurations, the increases in bulk temperature and enhanced heat transfer efficiency indicate that stress-free boundaries exhibit superior thermal transfer performance compared to no-slip boundaries. Consistent with this trend, heat transfer efficiency is further enhanced as both boundary surfaces become stress-free (INON), as shown in Figure \ref{fig: Tslices_type}(d). A steady and orderly laminar flow is built for high-efficiency heat transfer, with $Nu$ much enhanced by about $300\%$ compared to INON case. This is consistent with the stable flow regime previously observed by researchers on planar RBCs. The presence of curvature does not alter the stability of the roll flow regime \citep{wen_steady_2020, wang_zonal_2020}, and no zonal flow is observed in the ISOS cases. However, when only the inner boundary is changed to stress-free while the outer boundary remains no-slip (ISON, fig. \ref{fig: Tslices_type}(c)), a strong system-wide zonal flow emerges. The resulting mean azimuthal motion is aligned with the imposed rotation and produces intense shear, which leads to a pronounced reduction in $Nu$, consistent with the heat-transfer suppression reported in the sheared and oscillatory CC configurations \citep{zhong_sheared_2023, zhong_centrifugal_2025}. Since both the origin and the key features of this zonal-flow state are central to the present work, they are examined in detail in the following sections.

\subsection{Global Statistics}

First, we examine the response of global statistics to variations in the boundary condition sets, including the Nusselt number as a measure of heat-transfer efficiency and the Reynolds number as a measure of flow intensity. The definitions of $Nu$ and $Re$ in the CC systems read \citep{jiang_supergravitational_2020, wang_statistics_2022, wang_effects_2022}:

\begin{equation}
		Nu=\frac{\sqrt{RaPr}\langle u_r\theta\rangle_{t,\varphi}-\partial_r\langle\theta\rangle_{t,\varphi}}{(rln(\eta))^{-1}}, 
		Re= \sqrt{Ra/Pr}\sqrt{\langle |\boldsymbol{u}|^2 \rangle_{t,V}},
\end{equation} 
where the angle brackets $\langle\cdot\rangle_{t,\varphi}$ denote the average over the time $t$ and azimuthal direction $\varphi$, and the angle brackets $\langle\cdot\rangle_{t,V}$ denote the average over the time and space. 

\begin{figure}
    \centering
    \includegraphics[width=0.7\linewidth]{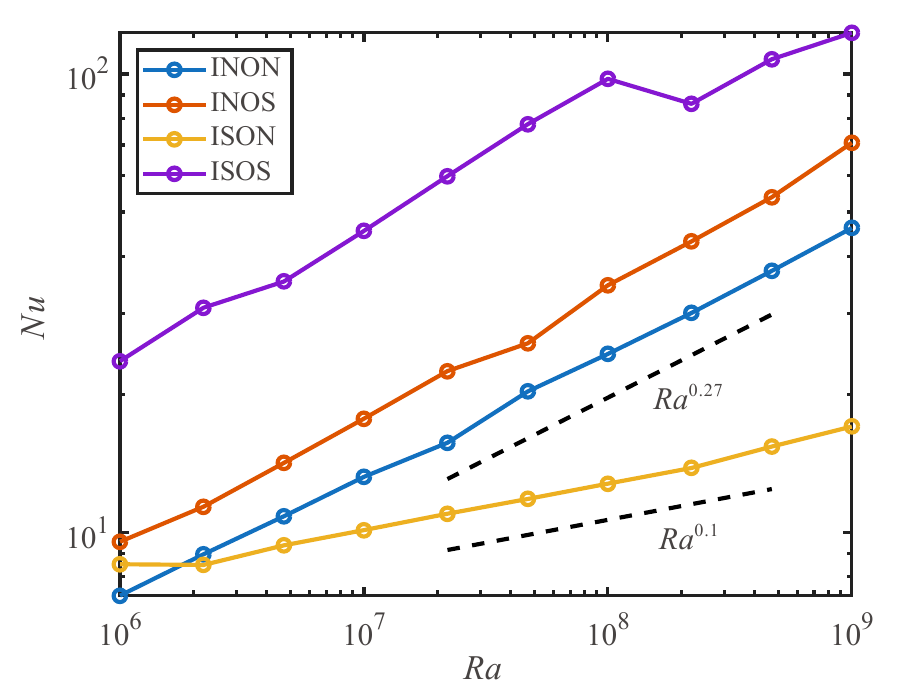}
    \captionsetup{justification=justified}
    \caption{ The variation of $Nu$ with $Ra$ at various boundary condition sets under fixed system geometry $\eta=0.5.$}
	\label{fig: Nu_Ra}
\end{figure}

Scaling relations of global statistics with $Ra$ often provide insight into the physical mechanisms underlying convective flows \citep{grossmann_scaling_2000, lohse_rmp_2024}. Accordingly, figure \ref{fig: Nu_Ra} presents $Nu$ as a function of $Ra\in [10^6, 10^9]$ for the different boundary-condition sets at the fixed geometry $\eta=0.5$. Under the same $Ra$, the heat transfer is typically strongest for ISOS, followed by INOS and then INON, while it is weakest for ISON, where a zonal flow develops. This ordering is consistent with the trends observed above, suggesting that it may reflect a relatively robust behavior. The only apparent exception occurs at $Ra=10^6$: under ISON conditions the zonal flow does not form, likely because the flow is too viscous and therefore insufficiently mobile, which results in a higher heat transfer than in the corresponding INON case. In addition, the slight but abrupt drop in heat transfer observed for the ISOS case at $Ra=2.2\times 10^8$ is associated with a change in the flow organisation from a four-vortex state to a two-vortex state, which may be linked to the stability of the convective-vortex aspect ratio under stress-free boundary conditions \citep{wang_zonal_2020}. 

Beyond this, the $Nu$--$Ra$ relations for the four boundary-condition sets clearly separate into two distinct scaling regimes. One regime follows $Nu\sim Ra^{0.27}$, in which the INON, INOS and ISON cases fall, and this scaling is consistent with the effective classical-regime behaviour commonly reported for RBC \citep{grossmann_scaling_2000, lohse_rmp_2024}. This suggests that, despite changes in the velocity boundary conditions, the dominant heat-transfer mechanism in the INOS case remains similar to that in the fully no-slip and fully stress-free configurations. In these cases, the LSC persists as the primary flow organisation, sweeping along the walls, extracting heat from the boundaries, and sustaining the development of the thermal boundary layers. In contrast, under ISON conditions where a zonal flow develops, the heat transfer follows a markedly different scaling, $Nu\sim Ra^{0.1}$. A similar reduction of the effective scaling exponent has also been reported for the zonal-flow state in planar RBC \citep{wang_zonal_2020}. Because the strong lateral shear associated with the zonal flow disrupts the coherent heat-transport pathway provided by the LSC, the overall heat transfer is substantially suppressed.

\begin{figure}
    \centering
    \includegraphics[width=1.0\linewidth]{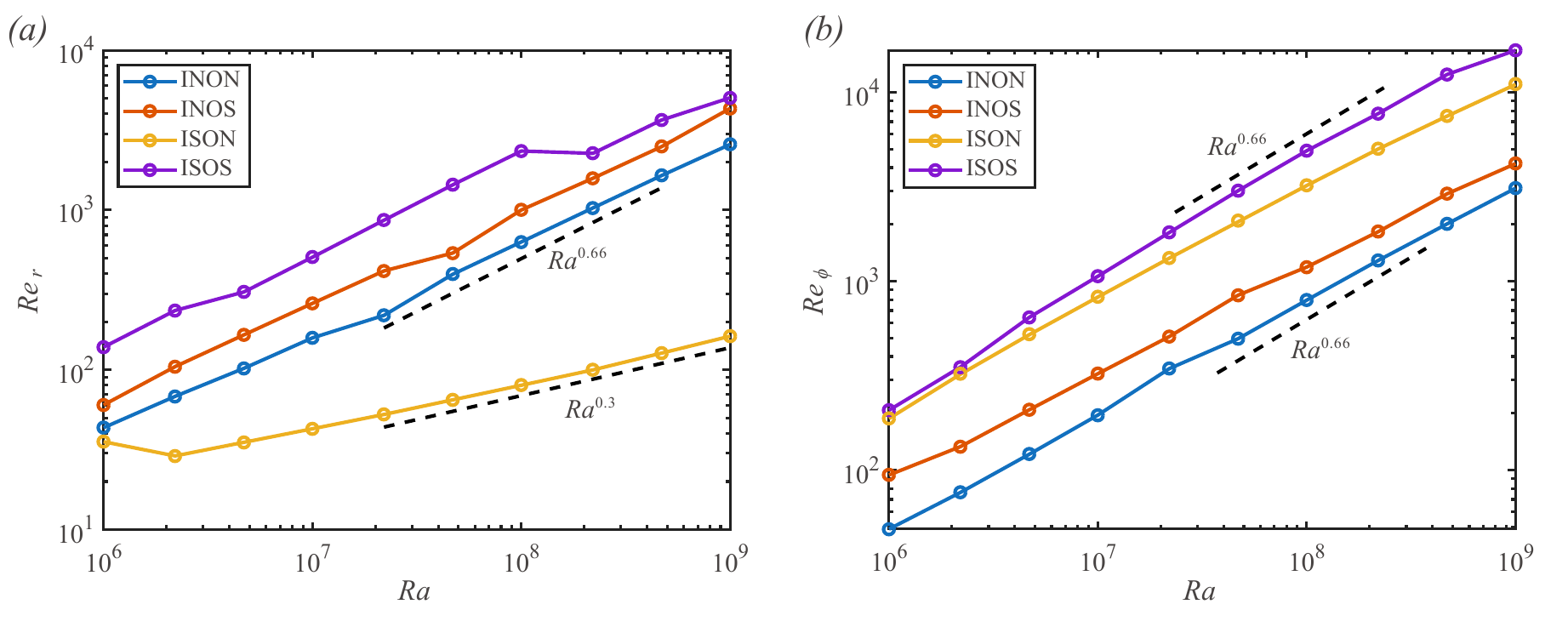}
    \captionsetup{justification=justified}
    \caption{ The variations of (a) the radial Reynolds number $Re_r$ and (b) the azimuthal Reynolds number $Re_\varphi$ with $Ra$ at various boundary condition sets under fixed system geometry $\eta=0.5.$}
	\label{fig: Re_T}
\end{figure}

Owing to the pronounced disparity between the azimuthal and radial motions in the zonal-flow state, we decompose the Reynolds number into directional components and analyse them separately; the results are presented in figure \ref{fig: Re_T}. For the INON, INOS and ISOS cases, the flow intensity follows the scaling $Re \sim Ra^{0.66}$ \citep{wen_steady_2020}, and their relative magnitudes are consistent with the corresponding ordering of the heat transfer. For the ISON case, the azimuthal Reynolds number exhibits a scaling similar to that of the other cases, although it remains weaker than in the ISOS configuration, whereas the radial Reynolds number is much smaller and follows $Re_r\sim Ra^{0.3}$. These results indicate that the emergence of zonal flow is accompanied by a strong anisotropy in the flow field.

\subsection{Time Evolution of the zonal flow} \label{subsec2}

In two-dimensional systems with stress-free boundaries, or in three-dimensional systems subject to strong rotational constraints, the flow can require long spin-up times because an inverse energy cascade transfers energy upscale. As a result, energy continues to accumulate on the largest scales until a statistically steady large-scale structure is established \citep{xu_fluctuation-induced_2024, kannan_beyond_2025}. Moreover, in two-dimensional RBC with stress-free plates, the zonal-flow state can be transient: it may ultimately decay into a roll state, with a survival probability that decreases exponentially in time \citep{wang_lifetimes_2023}, and the number of rolls may also evolve during this process. Therefore, tracking the temporal evolution of zonal-flow formation and decay is essential for clarifying its generation mechanism.

\begin{figure}
    \centering
    \includegraphics[width=0.8\linewidth]{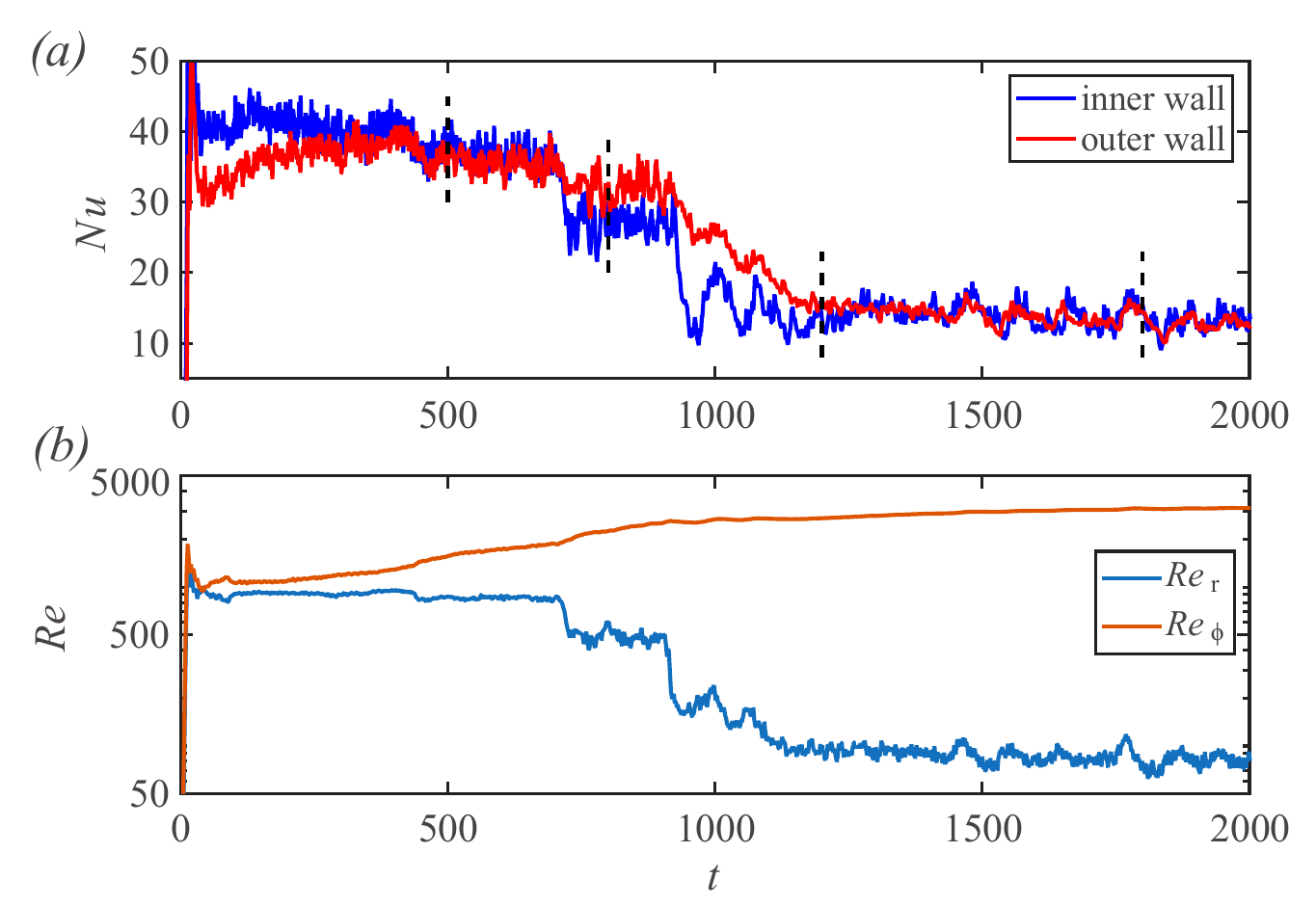}
    \captionsetup{justification=justified}
    \caption{ Time series of the (a) Nusselt number and (b) Reynolds number under the boundary condition set ISON, $Ra=10^8, \eta=0.5$. The dashed black lines in (a) indicate the sampling time for the snapshots shown in the figure \ref{fig: snapT}. }
	\label{fig: NuReT}
\end{figure}

\begin{figure}
    \centering
    \includegraphics[width=1.0\linewidth]{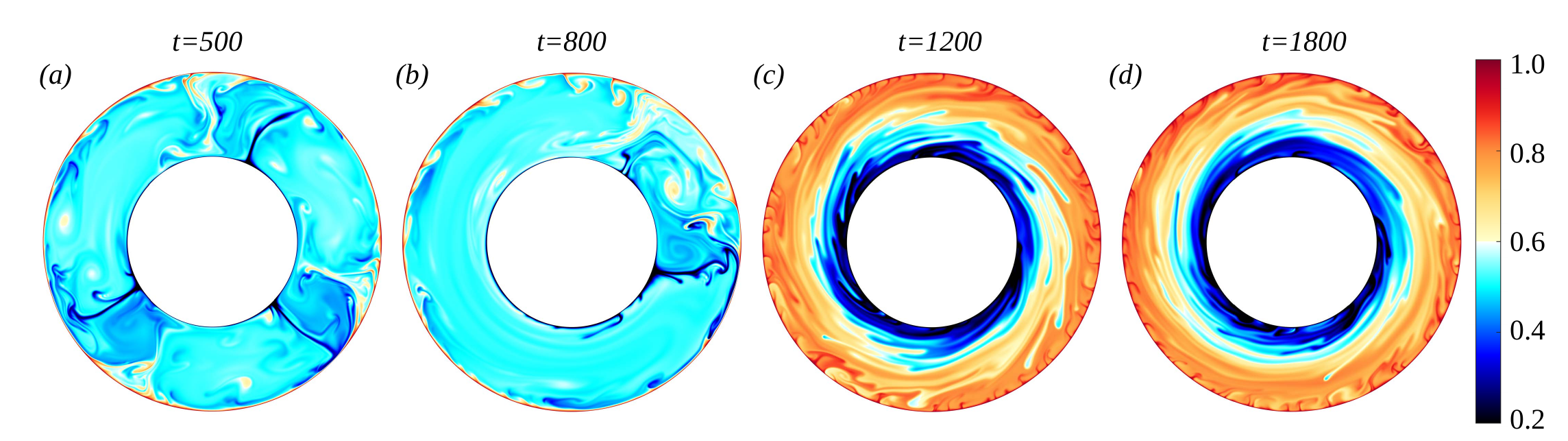}
    \captionsetup{justification=justified}
    \caption{ Typical instantaneous temperature snapshots and corresponding instantaneous Nusselt number under the boundary condition set ISON of different zonal flow evolution stages: (a) $t=500$, (b) $t=800$, (c) $t=1200$, (d) $t=1800$. $Ra=10^8, \eta=0.5.$}
	\label{fig: snapT}
\end{figure}

Figure \ref{fig: NuReT} shows the temporal evolution of the Nusselt number and Reynolds number for the ISON boundary-condition set at $Ra=10^8$ and $\eta=0.5$. Over the time window displayed, the simulation is initiated from random perturbations and finally evolves into the zonal-flow state. In the time traces of $Nu$ and $Re$, $Re_\varphi$ increases continuously, whereas $Re_r$ and $Nu$ decrease, showing a pronounced step-like decay. At the start of the simulation, the initially unstable flow adjusts rapidly and reaches an initial quasi-steady state after a few hundred dimensionless time units. In this stage, the heat transfer remains relatively strong, comparable to that in the INOS case ($Nu=34.6$), and the Reynolds numbers in both the radial and azimuthal directions are also of similar magnitude. An instantaneous snapshot of the temperature field at $t=500$ is shown in figure \ref{fig: snapT}(a). The LSC is clearly visible, with three pairs of convection rolls transporting heat between the inner and outer walls. At this stage, however, the flow is only quasi-steady: while the thermal fluid continuously recirculates within the rolls, kinetic energy gradually accumulates in the transverse zonal component, and the associated shear strengthens. This process persists until the zonal velocity exceeds a critical level, beyond which the original roll pattern can no longer be sustained, and the system transitions to the next stage.

Another instantaneous snapshot of the temperature field at $t=800$ is shown in figure \ref{fig: snapT}(b). By this time, the zonal flow has strengthened to nearly an order of magnitude larger than the radial flow. As the shear intensifies, the number of convection vortices decreases further, leaving only one roll pair, consisting of one small vortex and one large vortex. With continued evolution, the zonal flow keeps strengthening until the remaining roll structure can no longer be sustained once again. A relatively long transitional stage then follows, during which $Nu$ and $Re_r$ exhibit downward fluctuations as the convection rolls gradually weaken and eventually disappear. The corresponding instantaneous temperature snapshot at $t=1200$ is shown in figure \ref{fig: snapT}(c), where the system has reached a zonal-flow state characterized by low $Nu$ and strong $Re_\varphi$. Although $Re_\varphi$ continues to increase for some time thereafter, the zonal-flow state persists until $Re_\varphi$ also saturates. As shown in figure \ref{fig: snapT}(d), the temperature field at $t=1800$ remains similar to that at $t=1200$, except that the temperature gradient becomes more uniform, consistent with the reduced heat-transfer efficiency.

Similar temporal-evolution sequences are observed for ISON over a range of $Ra$, although the number of stages and their durations can differ from case to case. Such variations are likely caused by differences in $Ra$ and by the randomness of the initial perturbations. Overall, these observations indicate that the onset of the zonal flow state results from a gradual accumulation of kinetic energy. As in other two-dimensional flows featuring an inverse energy cascade, this build-up occurs on a viscous timescale, which is much longer than the convective timescale \citep{xu_fluctuation-induced_2024, kannan_beyond_2025}. During this slow upscale-transfer process, the system can pass through multiple metastable, convection-dominated states. Notably, these transient convective states closely resemble the flow patterns reported in the sheared CC system at different imposed shear levels \citep{zhong_sheared_2023}.

\subsection{Dissipation Analysis}

We further investigate the evolution of energy relationships within the system under varying flow conditions and flow regime transitions. Based on the energy balance built on classical RBC, the dimensionless time-dependence extension to the CC system can be derived for the equation \eqref{eq: OB} \citep{eckhardt_torque_2007, wang_effects_2022, zhong_sheared_2023}:

\begin{equation}\label{NuEqu}
    \frac{d\langle |\boldsymbol{u}|^2/2\rangle_V}{dt}=\frac{2(1-\eta)}{1+\eta}\langle ru_r\theta\rangle_V-\langle\varepsilon_u\rangle_V,
\end{equation}
where $\varepsilon_u=\frac{1}{2}\sqrt{\frac{Pr}{Ra}}\sum_{i,j}(\partial_iu_j+\partial_ju_i)^2$ is the dimensionless kinetic energy dissipation rate. When the energy supplied by buoyancy exceeds viscous dissipation, the surplus is converted into kinetic energy, thereby driving the further growth of the zonal flow. As a result, under ISON conditions with an established zonal flow, the system can remain in a long-lasting nonequilibrium state. Examining the spatial distribution of the kinetic-energy dissipation rate is therefore essential for clarifying the influence of the boundaries and the evolution of the zonal flow.

\begin{figure}
    \centering
    \includegraphics[width=1.0\linewidth]{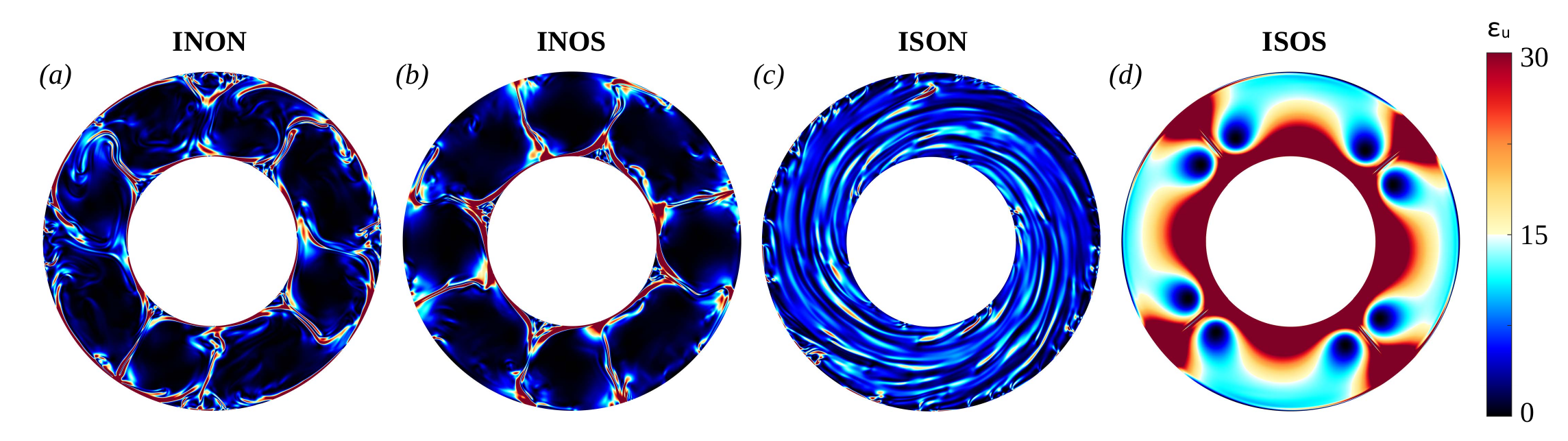}
    \captionsetup{justification=justified}
    \caption{ Typical instantaneous kinetic dissipation rate snapshots of different boundary condition sets: (a) INON, (b) INOS, (c) ISON, (d) ISOS. $Ra=10^8, \eta=0.5.$}
	\label{fig: dissip_type}
\end{figure}

Figure \ref{fig: dissip_type} presents representative instantaneous snapshots of the kinetic-energy dissipation rate for the four boundary-condition sets at fixed system geometry. From these distributions, the distinct flow responses to the different boundary conditions can be clearly identified. In the INON case, dissipation is primarily concentrated in the plumes and in the two thermal boundary layers. Under INOS conditions,  due to the stress-free conditions on the outer boundary, the dissipation is more concentrated at the inner boundary, while the LSC structure is still preserved. In the ISON case, once a zonal flow develops, a relatively low but spatially uniform dissipation level extends throughout the bulk, and no strong dissipation concentration is observed near the outer wall despite the no-slip boundary. The flow transitions from boundary layer-dominated to bulk-region-dominated. Finally, for ISOS, a thick high-dissipation region forms along the inner boundary, whereas at the outer boundary, only the areas swept by the plume motions exhibit noticeably enhanced dissipation.

From the comparison above, it is apparent that the inner and outer walls in the CC system bear different levels of dissipation. Because the inner wall has a smaller surface area, it is swept by more plumes per unit area, which can lead to a higher local dissipation intensity. For example, under ISOS conditions, dissipation near the inner wall is markedly stronger than near the outer wall. Under INOS conditions, the LSC can still be maintained; however, when the inner wall becomes no-slip in the ISON configuration, the LSC is much harder to sustain. Additionally, the curvature contrast between the inner and outer walls may also play an important role. Under the Coriolis force, plume streams emitted in the radial direction are deflected, with the corresponding deflection radius set by $Ro^{-1}$ \citep{wang_effects_2022}. Because the inner wall is outwardly convex (negative curvature), it is less effective at intercepting these deflected plumes than the inwardly concave (positive curvature) outer wall. Put differently, plumes emitted from the outer wall toward the inner wall are more prone to turn and travel along the inner boundary, which can preferentially feed kinetic energy into the zonal flow. 

\begin{figure}
    \centering
    \includegraphics[width=1.0\linewidth]{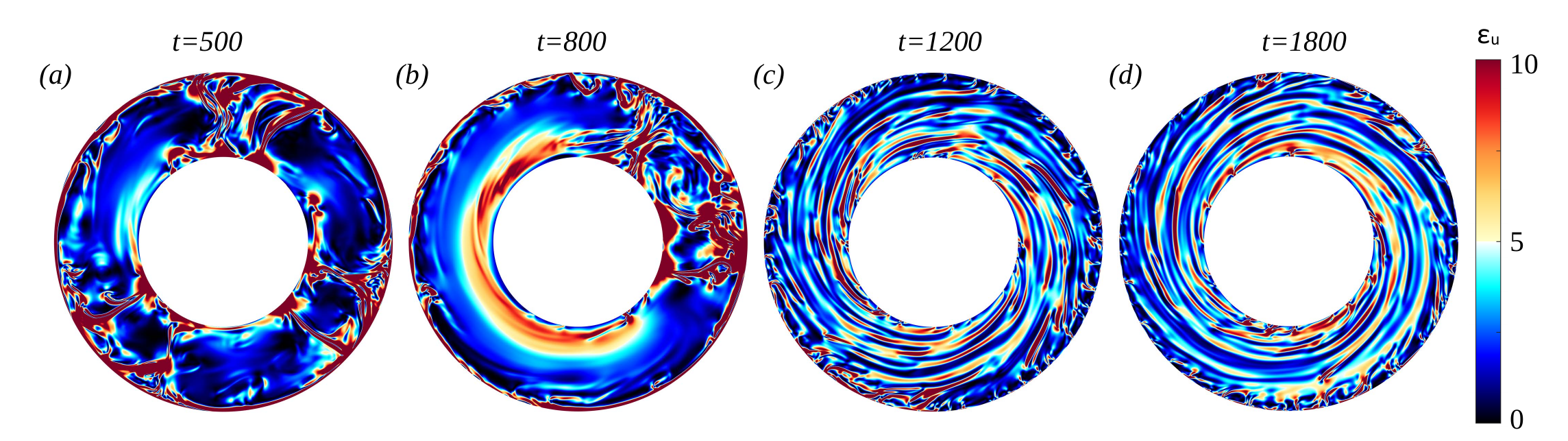}
    \captionsetup{justification=justified}
    \caption{ Typical instantaneous kinetic dissipation rate snapshots under the boundary condition set ISON of different zonal flow evolution stages: (a) $t=500$, (b) $t=800$, (c) $t=1200$, (d) $t=1800$. $Ra=10^8, \eta=0.5.$}
	\label{fig: dissip_time}
\end{figure}

Moreover, we examine how dissipation evolves during the formation of the zonal flow. Representative instantaneous snapshots of the kinetic-energy dissipation rate for the ISON boundary-condition set at different stages of the transition are shown in figure \ref{fig: dissip_time} (with a reduced colour range). Overall, the dissipation pattern in the bulk is consistent with the flow structures shown in figure \ref{fig: snapT}. Before the zonal flow is established, while the LSC remains present, dissipation is still strongly concentrated near the boundaries; even though the inner wall is stress-free, its dissipation level remains higher than that in the bulk, as illustrated in figure \ref{fig: dissip_time}(a). As the zonal flow strengthens, the large-scale convective organisation partially breaks down, and the dissipation field separates into two distinct regions: in the convection-dominated zone with coherent vortical structures, dissipation remains concentrated in plumes and boundary layers; in the extended region dominated by the zonal flow, plume activity is suppressed and boundary-layer dissipation is markedly reduced, as shown in figure \ref{fig: dissip_time}(b). As the zonal flow further intensifies, the high-dissipation region initially confined to the boundary layer expands into the bulk and reorganises into a counterclockwise banded pattern, as illustrated in figures \ref{fig: dissip_time}(c)(d).

\begin{figure}
    \centering
    \includegraphics[width=1.0\linewidth]{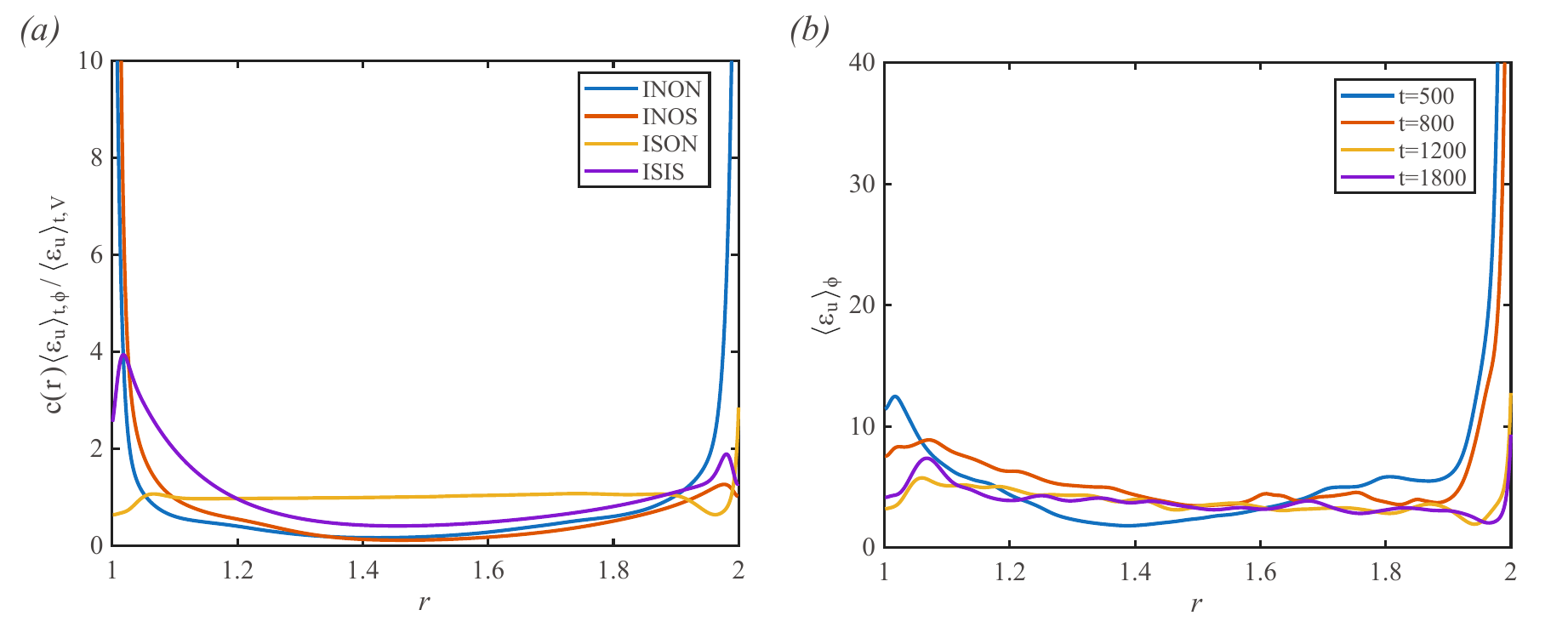}
    \captionsetup{justification=justified}
    \caption{ (a) Normalised area-weighted radial distribution of averaged kinetic dissipation rate under four different types of boundary condition sets. (b) Azimuthally averaged radial distribution of kinetic dissipation rate under the boundary condition set ISON of different zonal flow evolution stages. $Ra=10^8, \eta=0.5$.}
	\label{fig: dissip_curve}
\end{figure}

Furthermore, we perform a quantitative analysis of the dissipation statistics for the scenarios discussed above: (i) the radial distribution of dissipation in statistically steady flows under the different boundary-condition sets, and (ii) the corresponding distribution at successive stages of zonal-flow development under ISON conditions. The resulting dissipation-rate profiles as functions of radius are shown in Figure \ref{fig: dissip_curve}(a) and Figure \ref{fig: dissip_curve}(b), respectively. In Figure \ref{fig: dissip_curve}(a), to facilitate a consistent comparison of how dissipation is partitioned between the boundary layers and the bulk across different cases, we plot an area-weighted, normalised profile defined as $f(r)=\langle\varepsilon_u\rangle_{t,\varphi}r/\int_{r_i}^{r_o}\langle\varepsilon_u\rangle_{t,\varphi}rdr$. By construction, $\int_{r_i}^{r_o}f(r)dr=1$ holds for all cases. The results show that dissipation in both INON and INOS is strongly localised near the boundaries, indicating a boundary-layer-dominated dissipation. By contrast, under ISOS the stress-free boundaries lead to a more evenly distributed dissipation, with two pronounced peaks appearing near the inner and outer walls. For the ISON cases with zonal flow, the dissipation rate displays an unexpectedly uniform distribution in the bulk, with the normalised profile remaining close to a flat, nearly constant level. Moving from the bulk toward the boundaries, the dissipation decreases near both the inner and outer walls. On the inner side, where the boundary is stress-free, the profile drops directly to a local minimum at the wall, whereas near the no-slip outer wall the profile shows a secondary upturn. Under this configuration, the no-slip outer boundary appears to contribute mainly to an additional near-wall increase in dissipation required to satisfy the velocity constraint.

In figure \ref{fig: dissip_curve}(b), the azimuthally averaged dissipation-rate profiles illustrate how the dissipation intensity evolves during the development of zonal flow. At the early stage ($t=500$), the dissipation curve closely resembles that in the INOS case: dissipation is concentrated near the no-slip boundary, while a weaker peak is also present near the stress-free boundary. As time proceeds, dissipation near both walls—most noticeably near the no-slip outer wall—decreases gradually, whereas the bulk dissipation becomes increasingly uniform and eventually reaches a steady profile. This evolution suggests that the boundary-localised, high-dissipation configuration in the ISON setup is not sustained as the zonal flow strengthens. Instead, dissipation near the boundaries relaxes over time, consistent with the concomitant reduction in heat-transfer efficiency under the overall energy balance.

Overall, the dissipation analysis reveals a clear link between boundary conditions, flow organisation, and heat-transfer efficiency: cases that sustain the LSC (INON/INOS/ISOS) exhibit predominantly boundary-localised dissipation, whereas once a zonal flow develops (ISON) the system relaxes toward a low, nearly uniform bulk dissipation and weakened boundary-layer dissipation. This redistribution indicates that the boundary-localised, high-dissipation (and high-$Nu$) configuration is progressively eroded as energy is transferred into the large-scale azimuthal motion, ultimately leading to a shear-dominated regime with suppressed plume activity and reduced heat transport. The asymmetry between the inner and outer walls—through both unequal wall loading and the different ability of the boundaries to intercept deflected plumes—appears to be central to this transition. Motivated by these findings, the next subsection isolates the role of curvature by varying the annular radius ratio, in order to clarify how geometric curvature modifies the dissipation partition and the stability of the zonal-flow state.

\subsection{Effect of Geometry Asymmetry}

Building on the previous analysis, we suggest that the curvature asymmetry between the inner and outer walls may be a key ingredient in promoting the zonal-flow state observed under ISON conditions. In this subsection, we therefore examine how varying the curvature modifies the flow organisation, with particular attention to the onset and subsequent evolution of the zonal-flow state. We also quantify how curvature affects the system’s pronounced asymmetry, as reflected in the bulk temperature distribution.

\begin{figure}
    \centering
    \includegraphics[width=1.0\linewidth]{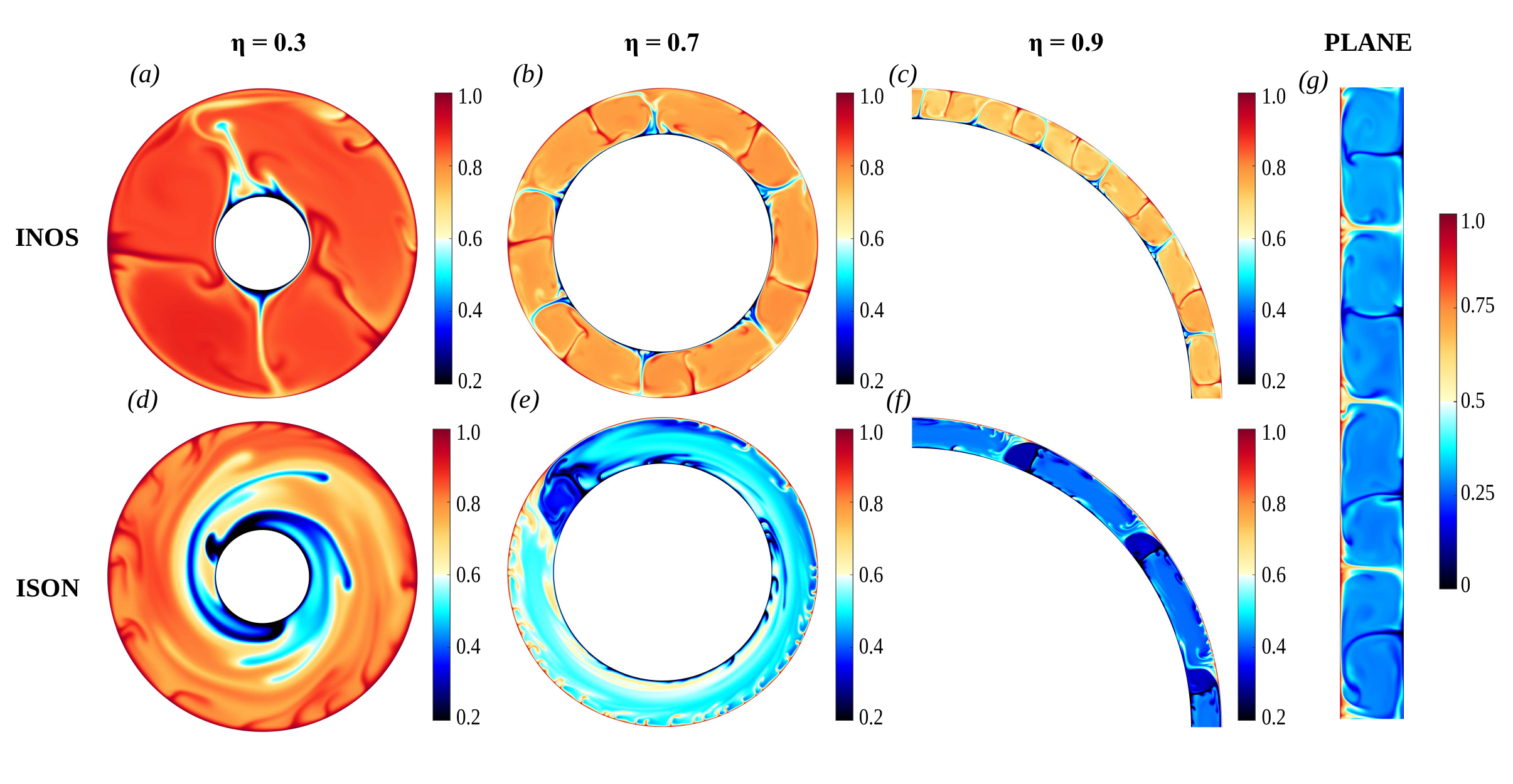}
    \captionsetup{justification=justified}
    \caption{ Typical instantaneous temperature snapshots under the boundary condition sets INOS and ISON of different radius ratios (a,d) $\eta=0.3$, (b, e) $\eta=0.7$, (c,f) $\eta=0.9$, (g) planar RBC. $Ra=10^7$.}
	\label{fig: snapEta}
\end{figure}

Figure \ref{fig: snapEta} summarises the final flow organisations under the two asymmetric boundary-condition sets (ISON and INOS) as the radius ratio $\eta$ increases from $0.3$ to $1$, where $\eta=1$ corresponds to the planar limit. For the INOS case, the LSC persists over the entire range of $\eta$, and the number of convection vortices increases with increasing $\eta$, consistent with previous observations under no-slip conditions \citep{wang_effects_2022, pitz_onset_2017}. Meanwhile, as $\eta$ increases, the asymmetry between the inner and outer walls weakens and the bulk temperature decreases gradually.

For the ISON case, the flow remains in a zonal-flow state at small $\eta$. When $\eta$ is increased to $0.7$, the system leaves the zonal-flow regime and develops a convection roll pair, whose structure resembles the intermediate stage shown in figure \ref{fig: snapT}(b). As $\eta$ increases further to $0.9$, multiple convection vortices appear, resembling the earlier stage illustrated in figure \ref{fig: snapT}(a). When $\eta$ reaches $1$ (the planar limit), the INOS and ISON cases converge to a common state characterised by regular Rayleigh--B\'enard convection rolls.

The results above indicate that, under ISON conditions, a robust zonal-flow state exhibits a greater tendency to occur at small radius ratios $\eta$, whereas increasing $\eta$ progressively drives the system away from the zonal-flow branch and back toward convection-roll states, with the planar limit ($\eta\to 1$) recovering regular Rayleigh-B\'enard rolls. This trend supports the view that curvature, and in particular the curvature asymmetry between the inner and outer walls, provides a geometric bias that promotes the formation and persistence of zonal flow under the Coriolis force. When $\eta$ is small, the strong curvature contrast implies that radially emitted plumes from the outer boundary, once deflected by rotation, can easily feed energy into the large-scale zonal component. As the zonal flow strengthens, the associated shear disrupts coherent roll-based transport, suppresses plume activity over extended regions, and is accompanied by a redistribution of the kinetic dissipation rate from a boundary-layer-dominated pattern toward a more uniform bulk-dominated field. Increasing $\eta$ reduces the curvature magnitude and weakens the geometric asymmetry, thereby diminishing this preferential pathway for azimuthal-momentum build-up and making the zonal-flow state harder to sustain. Consistent with this interpretation, at intermediate $\eta$ the system exhibits mixed or transitional roll structures reminiscent of the intermediate/early stages observed during zonal-flow development, before eventually settling into a convection-dominated organisation at larger $\eta$.

\begin{figure}
    \centering
    \includegraphics[width=1.0\linewidth]{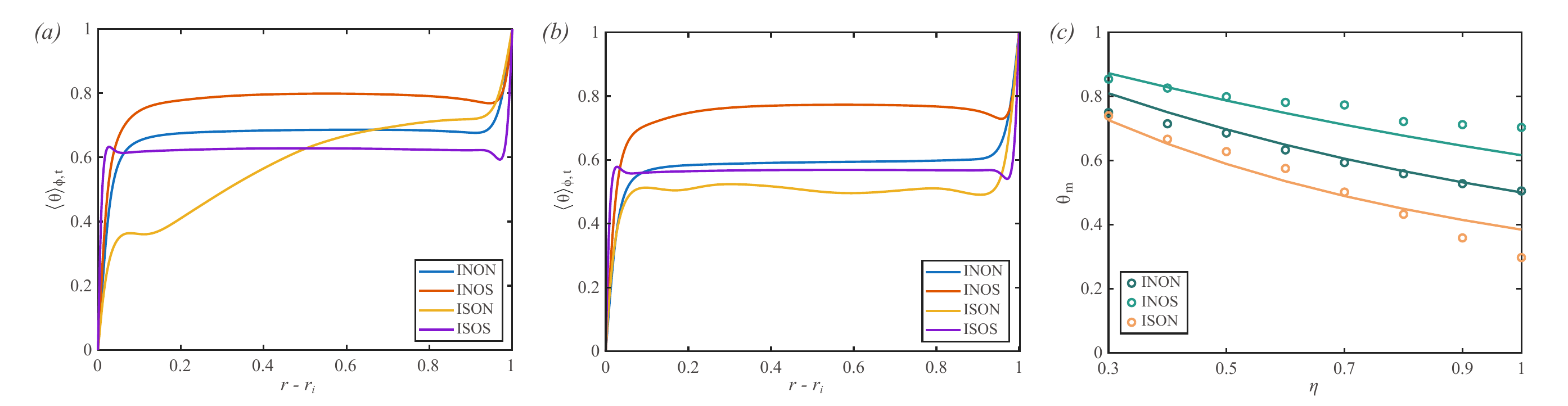}
    \captionsetup{justification=justified}
    \caption{ Radial distributions of temporally and azimuthally averaged temperature profiles of different boundary condition sets under $Ra=10^7$ and (a) $\eta=0.5$ (b) $\eta=0.7$. (c) Bulk temperature $\theta_m$ varies with the radius ratio $\eta$ of different boundary condition sets under $Ra=10^7$. The solid lines correspond to the theoretical results given by the convective laminar boundary layer model.}
	\label{fig: profile}
\end{figure}

Moreover, curvature asymmetry and boundary-condition asymmetry jointly shape the asymmetry of the temperature profiles. As shown in figure \ref{fig: Tslices_type} and figure \ref{fig: snapEta}, the bulk temperature varies markedly with both the boundary-condition set and the radius ratio. The radial profiles of the azimuthal- and time-averaged temperature at $\eta=0.5$ and $\eta=0.7$ are shown in figures \ref{fig: profile}(a) and (b), respectively. For fixed control parameters, the mean bulk temperature decreases in the order INOS, INON, ISOS, and ISON. In the LSC-dominated cases (INON, INOS, and ISOS), vigorous mixing in the bulk leads to an almost uniform bulk temperature. By contrast, under ISON at $\eta=0.5$, the system enters a zonal-flow state, mixing is weakened, and a pronounced bulk temperature gradient develops. When $\eta$ increases to $0.7$, the flow leaves the zonal-flow regime; enhanced convective mixing then restores a relatively uniform bulk temperature.

To quantify the bulk-temperature asymmetry, we follow the discussion of temperature-asymmetry models in RBC by \cite{fu_boundary_2025} and adopt the steady free-convective boundary-layer theory \citep{stewartson_free_1958, rotem_natural_1969}, which has been shown to perform well for cylindrical geometries. This framework describes buoyancy-driven convection developing above (or below) a heated (or cooled) semi-infinite plate with a single leading edge. Further details of the theoretical formulation are provided in Appendix \ref{app1}.

The dependence of the bulk temperature $\theta_m$ on the radius ratio $\eta$, together with the corresponding model predictions, is shown in figure \ref{fig: profile}(c). As $\eta$ increases, the bulk temperature decreases for all boundary-condition sets, while the relative ordering of $\theta_m$ among the cases remains unchanged. The steady free-convective boundary-layer theory captures both the observed trend with $\eta$ and the ordering between cases, showing good agreement with the DNS results. In this framework, the bulk temperature is set by the balance of heat fluxes between the inner and outer walls. The model further suggests that a stress-free wall transfers heat more efficiently than a no-slip wall; combined with the unequal surface areas of the inner and outer boundaries, this leads to the systematic shift of $\theta_m$ in the bulk.

\section{Conclusion} \label{sec4}

In the present study, we investigate rapidly rotating annular centrifugal convection with mixed (asymmetric) velocity boundary conditions, and quantify how boundary conditions and curvature jointly shape the flow organisation and heat transfer. Motivated by the quasi-two-dimensionalisation under strong rotation and the long integration times required to capture the zonal-flow dynamics, the parameter survey is carried out using two-dimensional direct numerical simulations. Four velocity boundary-condition sets are considered, including the symmetric cases (both no-slip and both stress-free) and the two asymmetric cases (one no-slip and one stress-free).

We find that the global transport properties depend strongly on the boundary condition set. In three cases, convection remains dominated by roll/LSC-type structures and the heat transfer exhibits an effective classical-type scaling behaviour with Rayleigh number, indicating a broadly similar transport mechanism despite the change in slip condition on one wall. In contrast, when the inner boundary is stress-free while the outer boundary remains no-slip (ISON), a strong system-wide zonal flow aligned with the imposed rotation emerges and leads to a pronounced suppression of heat transfer, accompanied by a much weaker effective $Nu$–$Ra$ scaling. Consistent with the flow reorganisation, the Reynolds number becomes strongly anisotropic in the zonal-flow state, with the azimuthal component remaining strong while the radial component is substantially reduced.

To clarify the underlying mechanism, we analyse the kinetic-energy dissipation field and its evolution during the transition into the zonal-flow state. While LSC and roll structures persist, dissipation remains strongly localised near the boundaries and in plume-rich regions, whereas the establishment of zonal flow coincides with a redistribution toward a relatively low and more uniform bulk dissipation and a marked weakening of boundary-layer dissipation. The time evolution further indicates that zonal-flow formation is a gradual energy-accumulation process with a long spin-up, during which the system can pass through multiple transient convection-dominated states before reaching the final zonal-flow regime.

Finally, by varying the radius ratio $\eta$, we show that curvature plays a key role in the emergence and persistence of zonal flow under asymmetric boundary conditions. In particular, increasing $\eta$ weakens the curvature asymmetry between the inner and outer walls and destabilises the zonal-flow branch, promoting a return to convection-roll states and leading to convergence toward the planar-limit behaviour as $\eta\to 1$. The accompanying bulk-temperature asymmetry varies systematically with both boundary conditions and curvature, and the observed trends can be interpreted in terms of the heat-flux balance between the two walls and their unequal geometric weighting.

In summary, the above results suggest that the zonal-flow branch in the present system is sustained by a delicate balance between boundary-condition asymmetry and geometric curvature, which together bias the flow toward azimuthal momentum build-up and weakened radial transport. An immediate next step is therefore to formulate quantitative criteria for the onset, persistence and decay of the zonal-flow state in terms of curvature and boundary asymmetry, and to link these criteria to measurable indicators such as $Re_\varphi/Re_r$ and the partition of dissipation between boundary layers and the bulk. It will also be important to test the robustness of the proposed mechanism in fully three-dimensional simulations and over broader ranges of control parameters (in particular rotation strength and Prandtl number), where additional instabilities and competing large-scale organisations may arise.

\backsection [Acknowledgements] {We thank Lei Ren, Rushi Lai and Jianjun Tao for insightful discussions.}

\backsection [Funding]{This work was supported by NSFC Excellence Research Group Program for ‘Multiscale Problems in Nonlinear Mechanics’ (No. 12588201), and the New Cornerstone Science Foundation through the New Cornerstone Investigator Program and the XPLORER PRIZE.}

\backsection[Declaration of interests] {The authors report no conflict of interest.}

\appendix
\section{Steady free-convective boundary-layer theory} \label{app1}

Steady free-convective boundary-layer (BL) theory provides a standard description for buoyancy-driven near-wall transport and is frequently adopted in Rayleigh-B\'enard-type analyses \citep{stewartson_free_1958, rotem_natural_1969}. In contrast to forced-convection BL models, the present framework does not prescribe an external ``edge'' velocity; instead, the pressure field adjusts so that its wall-normal gradient counteracts buoyancy while its wall-parallel gradient drives the tangential motion within the BL. As a result, the tangential velocity decays to zero outside the BL, consistent with a purely buoyancy-driven outer region.


Using local coordinates where $x$ is tangential to the wall and $z$ is measured normally outward from the wall into the fluid, the steady BL equations read
\begin{align}
\partial_x u_x + \partial_z u_z &= 0,
\label{eq:eq1_convective_BL_model}\\
u_x \partial_x u_x + u_z \partial_z u_x &= -\frac{1}{\rho}\partial_x p + \nu\,\partial_{zz} u_x,
\label{eq:eq2_convective_BL_model}\\
\frac{1}{\rho}\partial_z p &= s\, g\,\frac{\Delta\rho}{\rho},
\label{eq:pressure_eq_convective_BL_model}\\
u_x \partial_x \theta + u_z \partial_z \theta &= \kappa\,\partial_{zz}\theta.
\label{eq:eq4_convective_BL_model}
\end{align}
Here $s=-1$ for the inner BL and $s=1$ for the outer BL, reflecting the opposite orientations of the effective gravity in the local coordinate system at the two boundaries.

The kinematic conditions under the no-slip boundary condition are
\begin{equation}
u_x(0)=0,\quad u_z(0)=0,\quad u_x(z\to\infty)=0,\quad \partial_z u_x(z\to\infty)=0,
\label{eq:BC_velocity_convective_BL_model}
\end{equation}
so that no-slip holds at the wall while the tangential velocity vanishes smoothly at the BL edge. Under stress-free boundary conditions, there are:

\begin{equation}
\partial_z u_x(0)=0,\quad u_z(0)=0,\quad u_x(z\to\infty)=0,\quad \partial_z u_x(z\to\infty)=0,
\label{eq:BC_velocity_convective_BL_model_sf}
\end{equation}

The thermal boundary conditions are taken to be identical to those specified previously for the inner ($\theta(0)=\theta_i$) and outer walls ($\theta(0)=\theta_o$). 

Under the Boussinesq approximation, \eqref{eq:pressure_eq_convective_BL_model} can be written explicitly in terms of temperature as
\begin{align}
\frac{1}{\rho}\partial_z p &= -\alpha g_i\,(\theta-\theta_i), 
\label{eq:force_balance_inner}\\
\frac{1}{\rho}\partial_z p &= \alpha g_o\,(\theta-\theta_o),
\label{eq:force_balance_outer}
\end{align}
for the inner and outer BLs, respectively. $g_i=\Omega^2r_i, g_o=\Omega^2r_o$ are the local effective gravity at the inner and outer BLs.

Following the classical construction \citep{stewartson_free_1958,rotem_natural_1969}, the BL equations admit a similarity form.
Using the stream function $\hat{\Psi}(x,z)$ defined as $u_x=\partial_z\hat{\Psi},u_z=-\partial_x\hat{\Psi}$, introduce
\begin{align}
\xi &= \left(\frac{Ra}{\Pr}\right)^{1/5}\frac{z}{L_p}\left(\frac{x}{L_p}\right)^{-2/5},
\label{eq:xi}\\
\Psi(\xi) &= \left(\frac{Ra}{\Pr}\right)^{-1/5}\frac{1}{\nu}\left(\frac{x}{L_p}\right)^{-3/5}\hat{\Psi}(x,z),
\label{eq:Psi}\\
F(\xi) &= \left(\frac{Ra}{\Pr}\right)^{-4/5}\frac{2L_p^2}{\rho\nu^2}\left(\frac{x}{L_p}\right)^{-2/5}p(x,z),
\label{eq:F_pressure_convective_BL}\\
\Theta(\xi) &= \frac{\theta-\theta(0)}{\Delta}.
\label{eq:Theta_convective_BL}
\end{align}
The parameter $L_p$ denotes the along-wall plume spacing and may differ between inner and outer BLs, but in the moderate $Pr$, taking $L_p^i=L_p^o$ is reasonable \citep{fu_boundary_2025}. Therefore, in our system, we assume the along-wall plume spacing is the same between inner and outer BLs. Meanwhile, the $Ra$ in the similarity form denotes the local value defined by the effective gravity, hence it hold different value in the inner and outer BLs. 

Substitution of \eqref{eq:xi}--\eqref{eq:Theta_convective_BL} into the momentum and temperature equations yields the similarity system
\begin{align}
5\Psi''' + 3\Psi\Psi'' - (\Psi')^2 &= F - \xi F',
\label{eq:similarity_eq1_convective_BL_model}\\
\Theta'' + \frac{3}{5}\Pr\,\Psi\,\Theta' &= 0.
\label{eq:similarity_eq2_convective_BL_model}
\end{align}
The pressure--buoyancy relations \eqref{eq:force_balance_inner}--\eqref{eq:force_balance_outer} provide the closure between $F$ and $\Theta$.
For the inner BL, one obtains
\begin{equation}
F' + 2\Theta=0,
\label{eq:similarity_eq3_inner_convective_BL_model}
\end{equation}
while for the outer BL
\begin{equation}
F' + 2(1-\Theta)=0.
\label{eq:similarity_eq3_outer_convective_BL_model}
\end{equation}
Equations \eqref{eq:similarity_eq1_convective_BL_model}--\eqref{eq:similarity_eq2_convective_BL_model} together with \eqref{eq:similarity_eq3_inner_convective_BL_model} (inner) or \eqref{eq:similarity_eq3_outer_convective_BL_model} (outer) define the complete similarity problem.

For the stream-function profile, the boundary conditions under the no-slip case read
\begin{equation}
\Psi(0)=0,\quad \Psi'(0)=0,\quad \Psi'(\infty)=0,\quad \Psi''(\infty)=0,
\label{eq:BC_similarity_convective_BL_model}
\end{equation}
and under stress-free conditions, there is:
\begin{equation}
\Psi(0)=0,\quad \Psi''(0)=0,\quad \Psi'(\infty)=0,\quad \Psi''(\infty)=0,
\label{eq:BC_similarity_convective_BL_model_sf}
\end{equation}

and the thermal boundary conditions for $\Theta(\xi)$ are:

\begin{align}
    \Theta(0)=0, &\Theta(\infty)=\theta_m;\\
    \Theta(0)=1, &\Theta(\infty)=1-\theta_m,
\end{align}
for the inner and outer BLs, respectively. 

The bulk temperature $\theta_m$ links the solution of the inner boundary layer and the outer boundary layer, which is also limited by the conserved total heat flux, shown as:
\begin{equation}
        r_{i} \left \langle  \kappa \left | \partial_z\theta\right |_{z=0}^{i} \right \rangle_{x}  = r_{o} \left \langle \kappa \left | \partial_z\theta \right |_{z=0}^{o} \right \rangle_{x}.
\end{equation}
In the similarity forms, the conservation gives:
\begin{equation}
    \eta \left ( \frac{g_{i}}{g_{o}} \right )^{\frac{1}{5}} \left | \frac{\mathrm{d} \Theta}{\mathrm{d} \xi} \right |_{\xi=0}^{i} = \left | \frac{\mathrm{d} \Theta}{\mathrm{d} \xi} \right |_{\xi=0}^{o},
    \label{eq:matching_condition_convective_BL_model}
\end{equation}
where the scripts $i$ and $o$ denote the corresponding values for the inner boundary layer and outer boundary layer, respectively. Combining the above equations \eqref{eq:similarity_eq1_convective_BL_model}-\eqref{eq:matching_condition_convective_BL_model}, the temperature distribution and the bulk temperature in each no-slip and stress-free boundary condition set can be resolved.

\section{Simulation data} \label{app2}
We implemented our solver by extending the open-source AFiD framework originally developed for classical Rayleigh–B\'enard convection \citep{van_der_poel_pencil_2015, jiang_experimental_2022}, and adapted it to the present configuration by incorporating Coriolis effects, gravity, and centrifugal buoyancy. The discretization employs a staggered mesh; spatial operators are evaluated with second-order centered differences, while time advancement uses a third-order Runge–Kutta scheme coupled with Crank–Nicolson treatment of the implicit terms. To maintain stability of the explicitly integrated contributions, we restrict the Courant–Friedrichs–Lewy number to at most $0.7$. Considering the zonal flow state may be metastable and turn into a roll state \citep{wang_zonal_2020, wang_lifetimes_2023}, sufficient simulation time (larger than 5000 time units) is ensured for the flow in the zonal flow state to ensure stability. After the flow reaches a statistically stationary state, the simulations are continued for sufficiently long times to achieve satisfactory statistical convergence. The statistic errors of the Nusselt number are guaranteed to be within $2\%$.

The parameters of the main simulations considered in this work are listed in Table \ref{tab:appendix_tab1} and Table \ref{tab:appendix_tab2}. In Table \ref{tab:appendix_tab1}, the radius ratio $\eta=0.5$ is fixed while in Table \ref{tab:appendix_tab1} $Ra$ is fixed at $10^7$. The columns in Table \ref{tab:appendix_tab1} from left to right indicate the number of the simulation, Rayleigh numbers $Ra$, boundary condition sets, the resolution in the radial and azimuthal direction $(Nr, N_\varphi)$, the response Nusselt number $Nu$, Reynolds number in the radial and azimuthal direction $Re_r, Re_\varphi$ and the posterior check on the maximum grid spacing $\Delta_g$ by the global Kolmogorov length $\eta_K=(\nu^3/\varepsilon)^{1/4}$. In the Table \ref{tab:appendix_tab2}, the columns from left to right indicate the number of the simulation, the radius ratios $\eta$, boundary condition sets, the resolution in the radial and azimuthal direction $(Nr, N_\varphi)$, the response Nusselt number $Nu$, Reynolds number in the radial and azimuthal direction $Re_r, Re_\varphi$, the bulk temperature $\theta_m$ and the posterior check on the maximum grid spacing $\Delta_g$ by the global Kolmogorov length $\eta_K$ \citep{wang_effects_2022}.

\begin{table}
  \begin{center}
\def~{\hphantom{0}}
    \begin{tabular}{ccccccccc}
	$No.$ & $Ra$ & $BCs$ & $N_r$ & $N_\varphi$ & $Nu$ & $Re_r$ & $Re_\varphi$ & $\Delta_g/\eta_K$ \\
	$1$  & $1\times 10^{6}$   & $\mathrm{INON}$ & $128$ & $1536$ & $7.28$  & $43.44$   & $48.77$    & $0.196$ \\
	$2$  & $2.2\times 10^{6}$ & $\mathrm{INON}$ & $128$ & $1536$ & $8.96$  & $68.22$   & $76.29$    & $0.253$ \\
	$3$  & $4.7\times 10^{6}$ & $\mathrm{INON}$ & $128$ & $1536$ & $10.84$ & $102.21$  & $121.53$   & $0.322$ \\
	$4$  & $1\times 10^{7}$   & $\mathrm{INON}$ & $128$ & $1536$ & $13.21$ & $158.44$  & $195.77$   & $0.411$ \\
	$5$  & $2.2\times 10^{7}$ & $\mathrm{INON}$ & $192$ & $2304$ & $15.68$ & $219.27$  & $345.05$   & $0.349$ \\
	$6$  & $4.7\times 10^{7}$ & $\mathrm{INON}$ & $192$ & $2304$ & $20.30$ & $396.58$  & $497.16$   & $0.452$ \\
	$7$  & $1\times 10^{8}$   & $\mathrm{INON}$ & $256$ & $3072$ & $24.50$ & $631.51$  & $794.99$   & $0.430$ \\
	$8$  & $2.2\times 10^{8}$ & $\mathrm{INON}$ & $384$ & $4608$ & $30.10$ & $1030.03$ & $1289.69$  & $0.368$ \\
	$9$  & $4.7\times 10^{8}$ & $\mathrm{INON}$ & $384$ & $4608$ & $37.19$ & $1645.47$ & $2013.47$  & $0.470$ \\
	$10$ & $1\times 10^{9}$   & $\mathrm{INON}$ & $512$ & $6144$ & $46.13$ & $2577.00$ & $3111.07$  & $0.450$ \\
	$11$ & $1\times 10^{6}$   & $\mathrm{INOS}$ & $128$ & $1536$ & $9.54$  & $60.29$   & $94.27$    & $0.211$ \\
	$12$ & $2.2\times 10^{6}$ & $\mathrm{INOS}$ & $128$ & $1536$ & $11.37$ & $104.69$  & $133.32$   & $0.270$ \\
	$13$ & $4.7\times 10^{6}$ & $\mathrm{INOS}$ & $128$ & $1536$ & $14.17$ & $165.61$  & $208.91$   & $0.347$ \\
	$14$ & $1\times 10^{7}$   & $\mathrm{INOS}$ & $128$ & $1536$ & $17.70$ & $260.56$  & $324.12$   & $0.444$ \\
	$15$ & $2.2\times 10^{7}$ & $\mathrm{INOS}$ & $192$ & $2304$ & $22.48$ & $416.03$  & $510.13$   & $0.384$ \\
	$16$ & $4.7\times 10^{7}$ & $\mathrm{INOS}$ & $192$ & $2304$ & $25.85$ & $538.46$  & $844.95$   & $0.482$ \\
	$17$ & $1\times 10^{8}$   & $\mathrm{INOS}$ & $256$ & $3072$ & $34.60$ & $1001.10$ & $1189.60$  & $0.470$ \\
	$18$ & $2.2\times 10^{8}$ & $\mathrm{INOS}$ & $384$ & $4608$ & $43.11$ & $1579.85$ & $1830.58$  & $0.404$ \\
	$19$ & $4.7\times 10^{8}$ & $\mathrm{INOS}$ & $384$ & $4608$ & $53.86$ & $2495.74$ & $2905.68$  & $0.517$ \\
	$20$ & $1\times 10^{9}$   & $\mathrm{INOS}$ & $512$ & $6144$ & $70.77$ & $4327.59$ & $4197.85$  & $0.502$ \\
	$21$ & $1\times 10^{6}$   & $\mathrm{ISON}$ & $128$ & $1536$ & $8.52$  & $35.47$   & $187.72$   & $0.205$ \\
	$22$ & $2.2\times 10^{6}$ & $\mathrm{ISON}$ & $128$ & $1536$ & $8.49$  & $28.92$   & $322.25$   & $0.249$ \\
	$23$ & $4.7\times 10^{6}$ & $\mathrm{ISON}$ & $128$ & $1536$ & $9.37$  & $35.18$   & $521.88$   & $0.309$ \\
	$24$ & $1\times 10^{7}$   & $\mathrm{ISON}$ & $128$ & $1536$ & $10.11$ & $42.75$   & $825.28$   & $0.382$ \\
	$25$ & $2.2\times 10^{7}$ & $\mathrm{ISON}$ & $192$ & $2304$ & $10.98$ & $52.54$   & $1326.39$  & $0.317$ \\
	$26$ & $4.7\times 10^{7}$ & $\mathrm{ISON}$ & $192$ & $2304$ & $11.82$ & $64.96$   & $2085.64$  & $0.391$ \\
	$27$ & $1\times 10^{8}$   & $\mathrm{ISON}$ & $256$ & $3072$ & $12.77$ & $80.06$   & $3214.45$  & $0.362$ \\
	$28$ & $2.2\times 10^{8}$ & $\mathrm{ISON}$ & $384$ & $4608$ & $13.81$ & $99.84$   & $5017.16$  & $0.300$ \\
	$29$ & $4.7\times 10^{8}$ & $\mathrm{ISON}$ & $384$ & $4608$ & $15.38$ & $127.27$  & $7491.94$  & $0.373$ \\
	$30$ & $1\times 10^{9}$   & $\mathrm{ISON}$ & $512$ & $6144$ & $17.01$ & $162.34$  & $11024.76$ & $0.347$ \\
	$31$ & $1\times 10^{6}$   & $\mathrm{ISOS}$ & $128$ & $1536$ & $23.63$ & $138.70$  & $207.73$   & $0.269$ \\
	$32$ & $2.2\times 10^{6}$ & $\mathrm{ISOS}$ & $128$ & $1536$ & $30.91$ & $235.06$  & $351.75$   & $0.352$ \\
	$33$ & $4.7\times 10^{6}$ & $\mathrm{ISOS}$ & $128$ & $1536$ & $35.26$ & $307.21$  & $644.65$   & $0.440$ \\
	$34$ & $1\times 10^{7}$   & $\mathrm{ISOS}$ & $144$ & $1536$ & $45.45$ & $506.84$  & $1063.09$  & $0.567$ \\
	$35$ & $2.2\times 10^{7}$ & $\mathrm{ISOS}$ & $192$ & $2304$ & $59.87$ & $863.94$  & $1813.21$  & $0.494$ \\
	$36$ & $4.7\times 10^{7}$ & $\mathrm{ISOS}$ & $256$ & $3072$ & $77.74$ & $1438.96$ & $3020.63$  & $0.479$ \\
	$37$ & $1\times 10^{8}$   & $\mathrm{ISOS}$ & $384$ & $4608$ & $97.60$ & $2340.14$ & $4912.42$  & $0.408$ \\
	$38$ & $2.2\times 10^{8}$ & $\mathrm{ISOS}$ & $512$ & $6144$ & $86.08$ & $2262.85$ & $7728.97$  & $0.361$ \\
	$39$ & $4.7\times 10^{8}$ & $\mathrm{ISOS}$ & $512$ & $6144$ & $107.70$& $3660.77$ & $12452.48$ & $0.462$ \\
	$40$ & $1\times 10^{9}$   & $\mathrm{ISOS}$ & $768$ & $9216$ & $122.97$& $5044.54$ & $16713.88$ & $0.385$ \\
    \end{tabular}
    \captionsetup{justification=raggedright}
    \caption{Numerical details for simulations at fixed parameters $\eta=0.5, Pr=4.3$ and $Ro^{-1}=20$.}
\label{tab:appendix_tab1}
  \end{center}
\end{table}

\begin{table}
  \begin{center}
\def~{\hphantom{0}}
    \begin{tabular}{cccccccccc}
	$No.$ & $\eta$ & $BCs$ & $N_r$ & $N_\varphi$ & $Nu$ & $Re_r$ & $Re_\varphi$ & $\theta_m$ & $\Delta_g/\eta_K$ \\
	$1$  & $0.3$ & $\mathrm{INON}$ & $144$ & $1296$ & $11.65$ & $143.30$ & $239.38$ & $0.75$ & $0.330$ \\
	$2$  & $0.4$ & $\mathrm{INON}$ & $144$ & $1536$ & $12.69$ & $156.81$ & $208.66$ & $0.71$ & $0.336$ \\
	$3$  & $0.5$ & $\mathrm{INON}$ & $144$ & $1536$ & $13.27$ & $158.45$ & $195.12$ & $0.69$ & $0.411$ \\
	$4$  & $0.6$ & $\mathrm{INON}$ & $144$ & $2304$ & $13.59$ & $151.96$ & $189.60$ & $0.63$ & $0.346$ \\
	$5$  & $0.7$ & $\mathrm{INON}$ & $144$ & $3072$ & $14.26$ & $164.10$ & $183.39$ & $0.59$ & $0.352$ \\
	$6$  & $0.8$ & $\mathrm{INON}$ & $144$ & $4608$ & $14.20$ & $159.97$ & $188.34$ & $0.56$ & $0.352$ \\
	$7$  & $0.9$ & $\mathrm{INON}$ & $144$ & $2304$ & $14.28$ & $153.83$ & $188.90$ & $0.53$ & $0.353$ \\
	$8$  & $1$   & $\mathrm{INON}$ & $144$ & $1440$ & $14.28$ & $153.94$ & $190.42$ & $0.50$ & $0.360$ \\
	$9$  & $0.3$ & $\mathrm{INOS}$ & $144$ & $1296$ & $15.68$ & $221.36$ & $321.24$ & $0.85$ & $0.358$ \\
	$10$ & $0.4$ & $\mathrm{INOS}$ & $144$ & $1536$ & $16.69$ & $250.82$ & $323.69$ & $0.83$ & $0.362$ \\
	$11$ & $0.5$ & $\mathrm{INOS}$ & $144$ & $1536$ & $17.70$ & $260.28$ & $329.85$ & $0.80$ & $0.444$ \\
	$12$ & $0.6$ & $\mathrm{INOS}$ & $144$ & $2304$ & $18.68$ & $256.84$ & $352.22$ & $0.78$ & $0.377$ \\
	$13$ & $0.7$ & $\mathrm{INOS}$ & $144$ & $3072$ & $19.64$ & $241.62$ & $407.20$ & $0.77$ & $0.383$ \\
	$14$ & $0.8$ & $\mathrm{INOS}$ & $144$ & $4608$ & $20.05$ & $244.19$ & $357.83$ & $0.72$ & $0.386$ \\
	$15$ & $0.9$ & $\mathrm{INOS}$ & $144$ & $2304$ & $20.71$ & $259.94$ & $339.94$ & $0.71$ & $0.389$ \\
	$16$ & $1$   & $\mathrm{INOS}$ & $144$ & $1440$ & $20.90$ & $256.34$ & $325.01$ & $0.70$ & $0.398$ \\
	$17$ & $0.3$ & $\mathrm{ISON}$ & $144$ & $1296$ & $9.13$  & $43.36$  & $623.19$ & $0.74$ & $0.308$ \\
	$18$ & $0.4$ & $\mathrm{ISON}$ & $144$ & $1536$ & $9.51$  & $41.60$  & $730.64$ & $0.67$ & $0.311$ \\
	$19$ & $0.5$ & $\mathrm{ISON}$ & $144$ & $1536$ & $10.09$ & $42.72$  & $824.72$ & $0.63$ & $0.381$ \\
	$20$ & $0.6$ & $\mathrm{ISON}$ & $144$ & $2304$ & $10.79$ & $44.63$  & $906.63$ & $0.57$ & $0.325$ \\
	$21$ & $0.7$ & $\mathrm{ISON}$ & $144$ & $3072$ & $16.98$ & $127.64$ & $1165.09$ & $0.50$ & $0.369$ \\
	$22$ & $0.8$ & $\mathrm{ISON}$ & $144$ & $4608$ & $19.18$ & $211.15$ & $738.24$ & $0.43$ & $0.381$ \\
	$23$ & $0.9$ & $\mathrm{ISON}$ & $144$ & $2304$ & $19.58$ & $223.71$ & $680.01$ & $0.36$ & $0.384$ \\
	$24$ & $1$   & $\mathrm{ISON}$ & $144$ & $1440$ & $20.90$ & $256.34$ & $325.01$ & $0.30$ & $0.398$ \\
	$25$ & $0.3$ & $\mathrm{ISOS}$ & $144$ & $1296$ & $37.30$ & $393.10$ & $945.29$ & $0.73$ & $0.448$ \\
	$26$ & $0.4$ & $\mathrm{ISOS}$ & $144$ & $1536$ & $48.20$ & $581.14$ & $983.59$ & $0.67$ & $0.477$ \\
	$27$ & $0.5$ & $\mathrm{ISOS}$ & $144$ & $1536$ & $45.45$ & $506.84$ & $1063.09$ & $0.64$ & $0.567$ \\
	$28$ & $0.6$ & $\mathrm{ISOS}$ & $144$ & $2304$ & $48.37$ & $562.41$ & $1080.61$ & $0.59$ & $0.482$ \\
	$29$ & $0.7$ & $\mathrm{ISOS}$ & $144$ & $3072$ & $47.77$ & $546.73$ & $1108.34$ & $0.57$ & $0.482$ \\
	$30$ & $0.8$ & $\mathrm{ISOS}$ & $144$ & $4608$ & $46.99$ & $529.69$ & $1126.65$ & $0.54$ & $0.481$ \\
	$31$ & $0.9$ & $\mathrm{ISOS}$ & $144$ & $2304$ & $46.11$ & $512.08$ & $1137.74$ & $0.52$ & $0.479$ \\
	$32$ & $1$   & $\mathrm{ISOS}$ & $144$ & $1440$ & $45.99$ & $510.79$ & $1139.77$ & $0.50$ & $0.488$ \\
    \end{tabular}
    \captionsetup{justification=raggedright}
    \caption{Numerical details for simulations at fixed parameters $Ra=10^7, Pr=4.3$ and $Ro^{-1}=20$. In the simulations of $\eta=0.9$, we have limited the computational domain to only one-quarter of a circle to conserve computational resources. The validity of this approach has been verified by previous research \citep{wang_effects_2022, lai_gravity_2025}. Moreover, $\eta=1$ corresponds to the planar RBC with aspect ratio $L_x/L_z=10$.}
\label{tab:appendix_tab2}
  \end{center}
\end{table}

\bibliographystyle{jfm}
\bibliography{jfm}

@article{bhadra_heat_2025,
	title = {Heat and momentum transfer in {Rayleigh}–{Bénard} convection within a two-dimensional annulus under radial gravity},
	volume = {241},
	issn = {00179310},
	url = {https://linkinghub.elsevier.com/retrieve/pii/S0017931025000444},
	doi = {10.1016/j.ijheatmasstransfer.2025.126703},
	urldate = {2025-03-21},
	journal = {Int. J. Heat Mass Tran.},
	author = {Bhadra, A. and Shishkina, O. and Zhu, X.},
	month = may,
	year = {2025},
	pages = {126703},
}

@article{stewartson_free_1958,
	title = {On the free convection from a horizontal plate},
	volume = {9},
	issn = {1420-9039},
	url = {https://doi.org/10.1007/BF02033031},
	doi = {10.1007/BF02033031},
	year = {1958},
	pages = {276--282},
	number = {3},
	journal= {Z. Angew. Math. Phys},
	author = {Stewartson, K.},
	date = {1958-09-01},
}

@article{rotem_natural_1969,
	title = {Natural convection above unconfined horizontal surfaces},
	volume = {39},
	rights = {https://www.cambridge.org/core/terms},
	issn = {0022-1120, 1469-7645},
	url = {https://www.cambridge.org/core/product/identifier/S0022112069002102/type/journal_article},
	doi = {10.1017/S0022112069002102},
	pages = {173--192},
	number = {1},
	journal = {J. Fluid Mech.},
	author = {Rotem, Z. and Claassen, L.},
	urldate = {2026-01-04},
	date = {1969-10-23},
	langid = {english},
	year = {1969}
}

@article{wicht2019advances,
  title={Advances in geodynamo modelling},
  author={Wicht, J. and Sanchez, S.},
  journal={Geophys. Astrophys. Fluid Dyn.},
  volume={113},
  number={1-2},
  pages={2--50},
  year={2019},
  publisher={Taylor \& Francis}
}

@article{xu_fluctuation-induced_2024,
	title = {Fluctuation-induced transitions in anisotropic two-dimensional turbulence},
	volume = {9},
	issn = {2469-990X},
	url = {https://link.aps.org/doi/10.1103/PhysRevFluids.9.064605},
	doi = {10.1103/PhysRevFluids.9.064605},
	pages = {064605},
	number = {6},
	journal = {Phys. Rev. Fluids},
	author = {Xu, L. and Van Kan, A. and Liu, C. and Knobloch, E.},
	urldate = {2025-03-04},
	date = {2024-06-27},
	langid = {english},
	year = {2024}
}

@article{kannan_beyond_2025,
	title = {Beyond Nusselt number: assessing Reynolds and length scalings in rotating convection under stress-free boundary conditions},
	volume = {1016},
	issn = {0022-1120, 1469-7645},
	url = {https://www.cambridge.org/core/product/identifier/S0022112025103844/type/journal_article},
	doi = {10.1017/jfm.2025.10384},
	shorttitle = {Beyond Nusselt number},
	pages = {A3},
	journal = {J. Fluid Mech.},
	author = {Kannan, V. and Song, J. and Shishkina, O. and Zhu, X.},
	urldate = {2025-12-28},
	date = {2025-08-10},
	langid = {english},
	year = {2025},
}

@article{hanasoge2016seismic,
  title={Seismic sounding of convection in the Sun},
  author={Hanasoge, S. and Gizon, L. and Sreenivasan, K. R.},
  journal={Annu. Rev. Fluid Mech.},
  volume={48},
  pages={191--217},
  year={2016},
  publisher={Annual Reviews}
}

@article{hartmann_toward_2024,
	title = {Toward Understanding Polar Heat Transport Enhancement in Subglacial Oceans on Icy Moons},
	volume = {51},
	issn = {0094-8276, 1944-8007},
	url = {https://agupubs.onlinelibrary.wiley.com/doi/10.1029/2023GL105401},
	doi = {10.1029/2023GL105401},
    year = {2024},
	pages = {e2023GL105401},
	number = {3},
	journal = {Geophysical Research Letters},
	author = {Hartmann, R. and Stevens, R. J. A. M. and Lohse, D. and Verzicco, R.},
	urldate = {2025-02-17},
	date = {2024-02-16},
}

@article{1929Tropical,
  title={Tropical Convection and the Energy Balance at the Top of the Atmosphere},
  author={ Hartmann, D. L.  and  Moy, L. A.  and  Fu, Q. },
  journal={J. Clim},
  volume={14},
  number={24},
  pages={4495-4511},
  year={1929},
}

@article{bhadra_boundary-layer_2024,
	title = {On the boundary-layer asymmetry in two-dimensional annular {Rayleigh}–{Bénard} convection subject to radial gravity},
	volume = {999},
	issn = {0022-1120, 1469-7645},
	url = {https://www.cambridge.org/core/product/identifier/S0022112024009959/type/journal_article},
	doi = {10.1017/jfm.2024.995},
	language = {en},
	urldate = {2025-03-21},
	journal = {J. Fluid Mech.},
	author = {Bhadra, A. and Shishkina, O. and Zhu, X.},
	month = nov,
	year = {2024},
	pages = {R1},
}

@article{lohse_rmp_2024,
  title = {Ultimate Rayleigh-B\'enard turbulence},
  author = {Lohse, D. and Shishkina, O.},
  journal = {Rev. Mod. Phys.},
  volume = {96},
  issue = {3},
  pages = {035001},
  numpages = {60},
  year = {2024},
  month = {Aug},
}

@article{zhong_effect_2024,
	title = {Effect of radius ratio on the sheared annular centrifugal turbulent convection},
	volume = {992},
	issn = {0022-1120, 1469-7645},
	url = {https://www.cambridge.org/core/product/identifier/S0022112024005433/type/journal_article},
	doi = {10.1017/jfm.2024.543},
	urldate = {2024-09-14},
	journal = {J. Fluid Mech.},
	author = {Zhong, J. and Li, J. and Sun, C.},
	month = aug,
	year = {2024},
	pages = {A16}
}

@article{liu2020rayleigh,
  title={From Rayleigh--B{\'e}nard convection to porous-media convection: how porosity affects heat transfer and flow structure},
  author={Liu, S. and Jiang, L. and Chong, K. L. and Zhu, X. and Wan, Z.-H. and Verzicco, R. and Stevens, R. J. A. M. and Lohse, D. and Sun, C.},
  journal={J. Fluid Mech.},
  volume={895},
  year={2020},
  publisher={Cambridge University Press}
}

@article{zhong_2023,
  title={On the thermal effect of porous material in porous media {Rayleigh}–{Bénard} convection},
  author={Zhong, J. and Liu, S. and Sun, C.},
  journal={Flow},
  volume={3},
  pages={E13},
  year={2023},
  publisher={Cambridge University Press}
}

@article{yao2025direct,
  title={Direct numerical simulations of centrifugal convection: from gravitational to centrifugal buoyancy dominance},
  author={Yao, Z. and Emran, M. S. and Teimurazov, A. and Shishkina, O.},
  journal={Int. J. Heat Mass Transfer},
  volume={236},
  pages={126314},
  year={2025},
  publisher={Elsevier}
}

@article{zhong_sheared_2023,
	title = {From sheared annular centrifugal {Rayleigh}–{Bénard} convection to radially heated {Taylor}–{Couette} flow: exploring the impact of buoyancy and shear on heat transfer and flow structure},
	volume = {972},
	issn = {0022-1120, 1469-7645},
	shorttitle = {From sheared annular centrifugal {Rayleigh}–{Bénard} convection to radially heated {Taylor}–{Couette} flow},
	url = {https://www.cambridge.org/core/product/identifier/S0022112023007309/type/journal_article},
	doi = {10.1017/jfm.2023.730},
	urldate = {2023-10-15},
	journal = {J. Fluid Mech.},
	author = {Zhong, J. and Wang, D. and Sun, C.},
	month = oct,
	year = {2023},
	pages = {A29},
}

@article{yang_periodically_2020,
	title = {Periodically {Modulated} {Thermal} {Convection}},
	volume = {125},
	issn = {0031-9007, 1079-7114},
	url = {https://link.aps.org/doi/10.1103/PhysRevLett.125.154502},
	doi = {10.1103/PhysRevLett.125.154502},
	number = {15},
	urldate = {2023-07-31},
	journal = {Phys. Rev. Lett.},
	author = {Yang, R. and Chong, K. L. and Wang, Q. and Verzicco, R. and Shishkina, O. and Lohse, D.},
	month = oct,
	year = {2020},
	pages = {154502}
}

@article{van_der_poel_pencil_2015,
	title = {A pencil distributed finite difference code for strongly turbulent wall-bounded flows},
	volume = {116},
	issn = {00457930},
	url = {https://linkinghub.elsevier.com/retrieve/pii/S0045793015001164},
	doi = {10.1016/j.compfluid.2015.04.007},
	language = {en},
	urldate = {2022-11-21},
	journal = {Comput. Fluids},
	author = {van der Poel, E. P. and Ostilla-Mónico, R. and Donners, J. and Verzicco, R.},
	month = aug,
	year = {2015},
	pages = {10--16}
}

@article{verzicco_finite-difference_1996,
	title = {A {Finite}-{Difference} {Scheme} for {Three}-{Dimensional} {Incompressible} {Flows} in {Cylindrical} {Coordinates}},
	volume = {123},
	issn = {00219991},
	url = {https://linkinghub.elsevier.com/retrieve/pii/S0021999196900339},
	doi = {10.1006/jcph.1996.0033},
	language = {en},
	number = {2},
	urldate = {2022-11-21},
	journal = {J. Comput. Phys.},
	author = {Verzicco, R. and Orlandi, P.},
	month = feb,
	year = {1996},
	pages = {402--414}
}

@article{zhu_afid-gpu_2018,
	title = {{AFiD}-{GPU}: {A} versatile {Navier}–{Stokes} solver for wall-bounded turbulent flows on {GPU} clusters},
	volume = {229},
	issn = {00104655},
	shorttitle = {{AFiD}-{GPU}},
	url = {https://linkinghub.elsevier.com/retrieve/pii/S0010465518300985},
	doi = {10.1016/j.cpc.2018.03.026},
	language = {en},
	urldate = {2022-11-21},
	journal = {Comput. Phys. Commun.},
	author = {Zhu, X. and Phillips, E. and Spandan, V. and Donners, J. and Ruetsch, G. and Romero, J. and Ostilla-Mónico, R. and Yang, Y. and Lohse, D. and Verzicco, R. and Fatica, M. and Stevens, R. J.A.M.},
	month = aug,
	year = {2018},
	pages = {199--210}
}

@article{jiang_experimental_2022,
	title = {Experimental {Evidence} for the {Existence} of the {Ultimate} {Regime} in {Rapidly} {Rotating} {Turbulent} {Thermal} {Convection}},
	volume = {129},
	issn = {0031-9007, 1079-7114},
	url = {https://link.aps.org/doi/10.1103/PhysRevLett.129.204502},
	doi = {10.1103/PhysRevLett.129.204502},
	language = {en},
	number = {20},
	urldate = {2022-11-16},
	journal = {Phys. Rev. Lett.},
	author = {Jiang, H. and Wang, D. and Liu, S. and Sun, C.},
	month = nov,
	year = {2022},
	pages = {204502}
}

@article{xia_current_2013,
	title = {Current trends and future directions in turbulent thermal convection},
	volume = {3},
	issn = {20950349},
	url = {https://linkinghub.elsevier.com/retrieve/pii/S2095034915302531},
	doi = {10.1063/2.1305201},
	language = {en},
	number = {5},
	urldate = {2022-11-15},
	journal = {Theor. Appl. Mech. Lett.},
	author = {Xia, K.-Q.},
	year = {2013},
	pages = {052001}
}

@article{grossmann_scaling_2000,
	title = {Scaling in thermal convection: a unifying theory},
	volume = {407},
	issn = {0022-1120, 1469-7645},
	shorttitle = {Scaling in thermal convection},
	url = {https://www.cambridge.org/core/product/identifier/S0022112099007545/type/journal_article},
	doi = {10.1017/S0022112099007545},
	language = {en},
	urldate = {2022-10-13},
	journal = {J. Fluid Mech.},
	author = {Grossmann, S. and Lohse, D.},
	month = mar,
	year = {2000},
	pages = {27--56}
}

@article{grossmann_thermal_2001,
	title = {Thermal {Convection} for {Large} {Prandtl} {Numbers}},
	volume = {86},
	issn = {0031-9007, 1079-7114},
	url = {https://link.aps.org/doi/10.1103/PhysRevLett.86.3316},
	doi = {10.1103/PhysRevLett.86.3316},
	language = {en},
	number = {15},
	urldate = {2022-10-13},
	journal = {Phys. Rev. Lett.},
	author = {Grossmann, S. and Lohse, D.},
	month = apr,
	year = {2001},
	pages = {3316--3319}
}

@article{lohse_small-scale_2010,
	title = {Small-{Scale} {Properties} of {Turbulent} {Rayleigh}-{Bénard} {Convection}},
	volume = {42},
	issn = {0066-4189, 1545-4479},
	url = {https://www.annualreviews.org/doi/10.1146/annurev.fluid.010908.165152},
	doi = {10.1146/annurev.fluid.010908.165152},
	language = {en},
	number = {1},
	urldate = {2022-11-15},
	journal = {Annu. Rev. Fluid Mech.},
	author = {Lohse, D. and Xia, K.-Q.},
	month = jan,
	year = {2010},
	pages = {335--364}
}

@article{ahlers_heat_2009,
	title = {Heat transfer and large scale dynamics in turbulent {Rayleigh}-{Bénard} convection},
	volume = {81},
	issn = {0034-6861, 1539-0756},
	url = {https://link.aps.org/doi/10.1103/RevModPhys.81.503},
	doi = {10.1103/RevModPhys.81.503},
	language = {en},
	number = {2},
	urldate = {2022-11-15},
	journal = {Rev. Mod. Phys.},
	author = {Ahlers, G. and Grossmann, S. and Lohse, D.},
	month = apr,
	year = {2009},
	pages = {503--537}
}

@article{chilla_new_2012,
	title = {New perspectives in turbulent {Rayleigh}-{Bénard} convection},
	volume = {35},
	issn = {1292-8941, 1292-895X},
	url = {http://link.springer.com/10.1140/epje/i2012-12058-1},
	doi = {10.1140/epje/i2012-12058-1},
	language = {en},
	number = {7},
	urldate = {2022-11-15},
	journal = {Eur. Phys. J. E Soft Matter},
	author = {Chillà, F. and Schumacher, J.},
	month = jul,
	year = {2012},
	pages = {58}
}

@article{leng_flow_2021,
	title = {Flow structures and heat transport in {Taylor}–{Couette} systems with axial temperature gradient},
	volume = {920},
	issn = {0022-1120, 1469-7645},
	url = {https://www.cambridge.org/core/product/identifier/S0022112021004304/type/journal_article},
	doi = {10.1017/jfm.2021.430},
	language = {en},
	urldate = {2022-09-22},
	journal = {J. Fluid Mech.},
	author = {Leng, X.-Y. and Krasnov, D. and Li, B.-W. and Zhong, J.-Q.},
	month = aug,
	year = {2021},
	pages = {A42}
}

@article{jiang_supergravitational_2020,
	title = {Supergravitational turbulent thermal convection},
	volume = {6},
	issn = {2375-2548},
	url = {https://www.science.org/doi/10.1126/sciadv.abb8676},
	doi = {10.1126/sciadv.abb8676},
	language = {en},
	number = {40},
	urldate = {2022-09-22},
	journal = {Sci. Adv.},
	author = {Jiang, H. and Zhu, X. and Wang, D. and Huisman, S. G. and Sun, C.},
	month = oct,
	year = {2020},
	pages = {eabb8676}
}

@article{song_scaling_2024,
	title = {Scaling regimes in rapidly rotating thermal convection at extreme {Rayleigh} numbers},
	volume = {984},
	issn = {0022-1120, 1469-7645},
	url = {https://www.cambridge.org/core/product/identifier/S0022112024002490/type/journal_article},
	journal = {J. Fluid Mech.},
	author = {Song, J. and Shishkina, O. and Zhu, X.},
	month = apr,
	year = {2024},
	pages = {A45},
}

@article{ecke_2023_turbulent,
  title={Turbulent rotating {R}ayleigh--{B}{\'e}nard convection},
  author={Ecke, Robert E and Shishkina, O.},
  journal={Annu. Rev. Fluid Mech.},
  volume={55},
  pages={603--638},
  year={2023},
  publisher={Annual Reviews}
}

@article{ostilla_optimal_2013,
	title = {Optimal {Taylor}–{Couette} flow: direct numerical simulations},
	volume = {719},
	issn = {0022-1120, 1469-7645},
	shorttitle = {Optimal {Taylor}–{Couette} flow},
	url = {https://www.cambridge.org/core/product/identifier/S0022112012005964/type/journal_article},
	doi = {10.1017/jfm.2012.596},
	language = {en},
	urldate = {2022-09-22},
	journal = {J. Fluid Mech.},
	author = {Ostilla, R. and Stevens, R. J. A. M. and Grossmann, S. and Verzicco, R. and Lohse, D.},
	month = mar,
	year = {2013},
	pages = {14--46}
}

@article{blass_flow_2020,
	title = {Flow organization and heat transfer in turbulent wall sheared thermal convection},
	volume = {897},
	issn = {0022-1120, 1469-7645},
	url = {https://www.cambridge.org/core/product/identifier/S002211202000378X/type/journal_article},
	doi = {10.1017/jfm.2020.378},
	language = {en},
	urldate = {2022-09-22},
	journal = {J. Fluid Mech.},
	author = {Blass, A. and Zhu, X. and Verzicco, R. and Lohse, D. and Stevens, R. J. A. M.},
	month = aug,
	year = {2020},
	pages = {A22}
}

@article{wang_zonal_2020,
	title = {From zonal flow to convection rolls in {Rayleigh–Bénard} convection with free-slip plates},
	volume = {905},
	year = {2020},
	issn = {0022-1120, 1469-7645},
	url = {https://www.cambridge.org/core/product/identifier/S0022112020007934/type/journal_article},
	doi = {10.1017/jfm.2020.793},
	pages = {A21},
	journal = {J. Fluid Mech.},
	author = {Wang, Q. and Chong, K. L. and Stevens, R. J. A. M. and Verzicco, R. and Lohse, D.},
	urldate = {2025-10-16},
	date = {2020-12-25},
}

@article{lai_gravity_2025,
	title = {Gravity drives the flow within the Stewartson layer in centrifugal convection},
	volume = {1022},
	issn = {0022-1120, 1469-7645},
	url = {https://www.cambridge.org/core/product/identifier/S0022112025108082/type/journal_article},
	doi = {10.1017/jfm.2025.10808},
	year = {2025},
	pages = {A43},
	journal = {J. Fluid Mech.},
	author = {Lai, R. and Zhong, J. and Sun, C.},
	urldate = {2025-11-28},
	date = {2025-11-10},
	langid = {english},
}

@article{yadav_deep_2020,
	title = {Deep convection–driven vortex formation on Jupiter and Saturn},
	volume = {6},
	rights = {https://creativecommons.org/licenses/by-nc/4.0/},
	issn = {2375-2548},
	url = {https://www.science.org/doi/10.1126/sciadv.abb9298},
	doi = {10.1126/sciadv.abb9298},
	pages = {eabb9298},
	number = {46},
	journal = {Sci. Adv.},
	author = {Yadav, Rakesh Kumar and Heimpel, Moritz and Bloxham, Jeremy},
	urldate = {2024-11-17},
	date = {2020-11-13},
	year = {2020},
}

@article{fan_scaling_2024,
	title = {Scaling behaviour of rotating convection in a spherical shell with different Prandtl numbers},
	volume = {998},
	issn = {0022-1120, 1469-7645},
	url = {https://www.cambridge.org/core/product/identifier/S0022112024008930/type/journal_article},
	doi = {10.1017/jfm.2024.893},
	pages = {A20},
	journal = {J. Fluid Mech.},
	author = {Fan, W. and Wang, Q. and Lin, Y.},
	urldate = {2025-02-20},
	date = {2024-11-10},
	langid = {english},
	year = {2024}
}

@article{fu_turbulent_2024,
	title = {Turbulent spherical {Rayleigh–Bénard} convection: radius ratio dependence},
	volume = {1000},
	issn = {0022-1120, 1469-7645},
	url = {https://www.cambridge.org/core/product/identifier/S0022112024009091/type/journal_article},
	doi = {10.1017/jfm.2024.909},
	shorttitle = {Turbulent spherical {Rayleigh–Bénard} convection},
	pages = {A41},
	journal = {J. Fluid Mech.},
	author = {Fu, Y. and Bader, S. H. and Song, J. and Zhu, X.},
	urldate = {2024-12-04},
	date = {2024-12-10},
	langid = {english},
	year = {2024},
}

@article{kunnen_structure_2013,
	title = {The structure of sidewall boundary layers in confined rotating {Rayleigh–Bénard} convection},
	volume = {727},
	rights = {https://www.cambridge.org/core/terms},
	issn = {0022-1120, 1469-7645},
	url = {https://www.cambridge.org/core/product/identifier/S0022112013002851/type/journal_article},
	doi = {10.1017/jfm.2013.285},
	year ={2013},
	pages = {509--532},
	journal = {J. Fluid Mech.},
	author = {Kunnen, R. P. J. and Clercx, H. J. H. and Van Heijst, G. J. F.},
	urldate = {2024-10-28},
	date = {2013-07-25},
	langid = {english},
}

@article{kunnen_role_2011,
	title = {The role of Stewartson and Ekman layers in turbulent rotating {Rayleigh–Bénard} convection},
	volume = {688},
	rights = {https://www.cambridge.org/core/terms},
	issn = {0022-1120, 1469-7645},
	url = {https://www.cambridge.org/core/product/identifier/S0022112011003831/type/journal_article},
	doi = {10.1017/jfm.2011.383},
	pages = {422--442},
	journal= {J. Fluid Mech.},
	author = {Kunnen, R. P. J. and Stevens, R. J. A. M. and Overkamp, J. and Sun, C. and Van Heijst, G. F. and Clercx, H. J. H.},
	urldate = {2024-03-31},
	date = {2011-12-10},
	year = {2011}
}

@article{zhong_centrifugal_2025,
	title = {Centrifugal convection with oscillating rotating velocity},
	volume = {1010},
	issn = {0022-1120, 1469-7645},
	url = {https://www.cambridge.org/core/product/identifier/S002211202500326X/type/journal_article},
	doi = {10.1017/jfm.2025.326},
	pages = {A17},
	journal = {J. Fluid Mech.},
	author = {Zhong, J. and Li, J. and Zhou, Q. and Sun, C.},
	urldate = {2025-06-18},
	date = {2025-05-10},
	langid = {english},
	year = {2025},
}

@article{yerragolam_scaling_2024,
	title = {Scaling relations for heat and momentum transport in sheared {Rayleigh–Bénard} convection},
	volume = {1000},
	issn = {0022-1120, 1469-7645},
	url = {https://www.cambridge.org/core/product/identifier/S0022112024008723/type/journal_article},
	doi = {10.1017/jfm.2024.872},
	pages = {A74},
	journal = {J. Fluid Mech.},
	author = {Yerragolam, G. S. and Howland, C. J. and Stevens, R. J.A.M. and Verzicco, R. and Shishkina, O. and Lohse, D.},
	urldate = {2025-01-21},
	date = {2024-12-10},
	langid = {english},
	year = {2024},
}

@article{song_direct_2024,
	title = {Direct numerical simulations of rapidly rotating {Rayleigh–Bénard} convection with Rayleigh number up to},
	volume = {989},
	issn = {0022-1120, 1469-7645},
	url = {https://www.cambridge.org/core/product/identifier/S0022112024004841/type/journal_article},
	doi = {10.1017/jfm.2024.484},
	pages = {A3},
	journal = {J. Fluid Mech.},
	author = {Song, J. and Shishkina, O. and Zhu, X.},
	urldate = {2025-02-16},
	date = {2024-06-25},
	year = {2024},
}

@article{li_darcy_2025,
	title = {From Darcy convection to free-fluid convection: pore-scale study of density-driven currents in porous media},
	volume = {1009},
	issn = {0022-1120, 1469-7645},
	url = {https://www.cambridge.org/core/product/identifier/S0022112025002137/type/journal_article},
	doi = {10.1017/jfm.2025.213},
	shorttitle = {From Darcy convection to free-fluid convection},
	pages = {A10},
	journal = {J. Fluid Mech.},
	author = {Li, J. and Yang, Y. and Sun, C.},
	urldate = {2025-12-21},
	date = {2025-04-25},
	langid = {english},
	year = 2025,
}

@article{zhao_modulation_2022,
	title = {Modulation of turbulent Rayleigh-Bénard convection under spatially harmonic heating},
	volume = {105},
	issn = {2470-0045, 2470-0053},
	url = {https://link.aps.org/doi/10.1103/PhysRevE.105.055107},
	doi = {10.1103/PhysRevE.105.055107},
	pages = {055107},
	number = {5},
	journal = {Phys. Rev. E},
	author = {Zhao, C.-B. and Zhang, Y.-Z. and Wang, B.-F. and Wu, J.-Z. and Chong, K. L. and Zhou, Q.},
	urldate = {2023-07-31},
	date = {2022-05-23},
	year = {2022},
}

@article{zhou_deep_2025,
	title = {Deep reinforcement learning control unlocks enhanced heat transfer in turbulent convection},
	volume = {122},
	issn = {0027-8424, 1091-6490},
	url = {https://pnas.org/doi/10.1073/pnas.2506351122},
	doi = {10.1073/pnas.2506351122},
	pages = {e2506351122},
	number = {37},
	journal = {Proc. Natl. Acad. Sci. U.S.A.},
	author = {Zhou, Z. and Zhu, X.},
	urldate = {2025-12-21},
	date = {2025-09-16},
	year={2025},
}

@article{van_der_poel_effect_2014,
	title = {Effect of velocity boundary conditions on the heat transfer and flow topology in two-dimensional {R}ayleigh-{B}énard convection},
	volume = {90},
	issn = {1539-3755, 1550-2376},
	url = {https://link.aps.org/doi/10.1103/PhysRevE.90.013017},
	doi = {10.1103/PhysRevE.90.013017},
	pages = {013017},
	number = {1},
	journal = {Phys. Rev. E},
	author = {Van Der Poel, E. P. and Ostilla-Mónico, R. and Verzicco, R. and Lohse, D.},
	urldate = {2023-06-08},
	date = {2014-07-23},
	langid = {english},
	year = {2014}
}

@article{fu_boundary_2025,
	title = {Boundary layer asymmetry in turbulent spherical {Rayleigh–Bénard} convection: combined dependence on Prandtl number and radius ratio},
	volume = {1018},
	issn = {0022-1120, 1469-7645},
	url = {https://www.cambridge.org/core/product/identifier/S0022112025104977/type/journal_article},
	doi = {10.1017/jfm.2025.10497},
	shorttitle = {Boundary layer asymmetry in turbulent spherical {Rayleigh–Bénard} convection},
	year = {2025}, 
	pages = {A21},
	journal= {J. Fluid Mech.},
	author = {Fu, Y. and Bader, S. H. and Zhu, X.},
	urldate = {2026-01-04},
	date = {2025-09-10},
	langid = {english},
}

@article{wen_steady_2020,
	title = {Steady {Rayleigh–Bénard} convection between stress-free boundaries},
	volume = {905},
	issn = {0022-1120, 1469-7645},
	url = {https://www.cambridge.org/core/product/identifier/S0022112020008125/type/journal_article},
	doi = {10.1017/jfm.2020.812},
	pages = {R4},
	journal = {J. Fluid Mech.},
	author = {Wen, B. and Goluskin, Da. and {LeDuc}, M.and Chini, G. P. and Doering, C. R.},
	urldate = {2025-09-26},
	date = {2020-12-25},
	langid = {english},
	year={2020},
}

@article{heimpel_polar_2022,
	title = {Polar and mid-latitude vortices and zonal flows on Jupiter and Saturn},
	volume = {379},
	issn = {00191035},
	url = {https://linkinghub.elsevier.com/retrieve/pii/S001910352200063X},
	doi = {10.1016/j.icarus.2022.114942},
	pages = {114942},
	journal= {Icarus},
	author = {Heimpel, M. H. and Yadav, R. K. and Featherstone, N. A. and Aurnou, J. M.},
	urldate = {2025-01-20},
	date = {2022-06},
	langid = {english},
	year = {2022},
}

@article{Schneider2006The,
   author = {Schneider, T.},
   title = {The General Circulation of the Atmosphere}, 
   journal= {Annu. Rev. Earth Planet. Sci},
   year = {2006},
   volume = {34},
   number = {Volume 34, 2006},
   pages = {655-688},
   doi = "https://doi.org/10.1146/annurev.earth.34.031405.125144",
   url = "https://www.annualreviews.org/content/journals/10.1146/annurev.earth.34.031405.125144",
   publisher = "Annual Reviews",
   issn = "1545-4495",
  }

@article{zhu_transport_2024,
	title = {Transport scaling in porous media convection},
	volume = {991},
	issn = {0022-1120, 1469-7645},
	url = {https://www.cambridge.org/core/product/identifier/S0022112024005287/type/journal_article},
	doi = {10.1017/jfm.2024.528},
	pages = {A4},
	journal = {J. Fluid Mech.},
	author = {Zhu, X. and Fu, Y. and De Paoli, M.},
	urldate = {2025-12-21},
	date = {2024-07-25},
	langid = {english},
	year = {2024},
}

@article{xu_pore-scale_2023,
	title = {Pore-scale statistics of temperature and thermal energy dissipation rate in turbulent porous convection},
	volume = {8},
	issn = {2469-990X},
	url = {https://link.aps.org/doi/10.1103/PhysRevFluids.8.093504},
	doi = {10.1103/PhysRevFluids.8.093504},
	pages = {093504},
	number = {9},
	journal = {Phys. Rev. Fluids},
	author = {Xu, A. and Xu, B.-R. and Xi, H.-D.},
	urldate = {2025-12-21},
	date = {2023-09-18},
	langid = {english},
	year={2023}
}

@article{wang_lifetimes_2023,
	title = {Lifetimes of metastable windy states in two-dimensional {Rayleigh–Bénard} convection with stress-free boundaries},
	volume = {976},
	issn = {0022-1120, 1469-7645},
	url = {https://www.cambridge.org/core/product/identifier/S0022112023008753/type/journal_article},
	doi = {10.1017/jfm.2023.875},
	year={2023},
	pages = {R2},
	journal = {J. Fluid Mech.},
	author = {Wang, Q. and Goluskin, D. and Lohse, D.},
	urldate = {2025-09-26},
	date = {2023-12-10},
	langid = {english},
}

@article{yeh_geophysical_2023,
	title = {Geophysical fluid dynamics in the hypergravity field},
	volume = {40},
	issn = {1614-3116},
	url = {https://doi.org/10.1007/s10409-023-23296-x},
	doi = {10.1007/s10409-023-23296-x},
	number = {2},
	journal = {Acta Mech. Sin.},
	author = {Yeh, H.},
	month = sep,
	year = {2023},
	pages = {723296},
}

@article{wang_statistics_2022,
	title = {Statistics of temperature and velocity fluctuations in supergravitational convective turbulence},
	volume = {39},
	issn = {1614-3116},
	url = {https://doi.org/10.1007/s10409-022-22387-x},
	doi = {10.1007/s10409-022-22387-x},
	number = {4},
	journal = {Acta Mech. Sin.},
	author = {Wang, D. and Liu, J. and Zhou, Q. and Sun, C.},
	month = dec,
	year = {2022},
	pages = {122387},
}

@article{king_boundary_2009,
	title = {Boundary layer control of rotating convection systems},
	volume = {457},
	copyright = {http://www.springer.com/tdm},
	issn = {0028-0836, 1476-4687},
	url = {https://www.nature.com/articles/nature07647},
	doi = {10.1038/nature07647},
	language = {en},
	number = {7227},
	urldate = {2026-01-26},
	journal = {Nature},
	author = {King, E. M. and Stellmach, S. and Noir, J. and Hansen, U. and Aurnou, J. M.},
	month = jan,
	year = {2009},
	pages = {301--304},
}

@article{king_turbulent_2013,
	title = {Turbulent convection in liquid metal with and without rotation},
	volume = {110},
	issn = {0027-8424, 1091-6490},
	url = {https://pnas.org/doi/full/10.1073/pnas.1217553110},
	doi = {10.1073/pnas.1217553110},
	number = {17},
	urldate = {2026-01-26},
	journal = {Proc. Natl. Acad. Sci. U.S.A.},
	author = {King, E. M. and Aurnou, J. M.},
	month = apr,
	year = {2013},
	pages = {6688--6693},
}

@article{wang_effects_2022,
	title = {Effects of radius ratio on annular centrifugal {Rayleigh}–{Bénard} convection},
	volume = {930},
	issn = {0022-1120, 1469-7645},
	url = {https://www.cambridge.org/core/product/identifier/S0022112021008892/type/journal_article},
	doi = {10.1017/jfm.2021.889},

	language = {en},
	urldate = {2022-09-22},
	journal = {J. Fluid Mech.},
	author = {Wang, D. and Jiang, H. and Liu, S. and Zhu, X. and Sun, C.},
	month = jan,
	year = {2022},
	pages = {A19}
}

@article{eckhardt_torque_2007,
	title = {Torque scaling in turbulent {Taylor}–{Couette} flow between independently rotating cylinders},
	volume = {581},
	issn = {0022-1120, 1469-7645},
	url = {https://www.cambridge.org/core/product/identifier/S0022112007005629/type/journal_article},
	doi = {10.1017/S0022112007005629},
	language = {en},
	urldate = {2022-09-23},
	journal = {J. Fluid Mech.},
	author = {Eckhardt, B. and Grossmann, S. and Lohse, D.},
	month = jun,
	year = {2007},
	pages = {221--250}
}

@article{pitz_onset_2017,
	title = {Onset of convection induced by centrifugal buoyancy in a rotating cavity},
	volume = {826},
	issn = {0022-1120, 1469-7645},
	url = {https://www.cambridge.org/core/product/identifier/S0022112017004517/type/journal_article},
	doi = {10.1017/jfm.2017.451},
	language = {en},
	urldate = {2022-10-04},
	journal = {J. Fluid Mech.},
	author = {Pitz, D. B. and Marxen, O. and Chew, J. W.},
	month = sep,
	year = {2017},
	pages = {484--502}
}

@article{goluskin_convectively_2014,
	title = {Convectively driven shear and decreased heat flux},
	volume = {759},
	issn = {0022-1120, 1469-7645},
	url = {https://www.cambridge.org/core/product/identifier/S0022112014005771/type/journal_article},
	doi = {10.1017/jfm.2014.577},
	language = {en},
	urldate = {2022-10-25},
	journal = {J. Fluid Mech.},
	author = {Goluskin, D. and Johnston, H. and Flierl, G. R. and Spiegel, E. A.},
	month = nov,
	year = {2014},
	pages = {360--385}
}

@article{rouhi_coriolis_2021,
	title = {Coriolis effect on centrifugal buoyancy-driven convection in a thin cylindrical shell},
	volume = {910},
	issn = {0022-1120, 1469-7645},
	url = {https://www.cambridge.org/core/product/identifier/S0022112020009593/type/journal_article},
	doi = {10.1017/jfm.2020.959},
	language = {en},
	urldate = {2022-11-08},
	journal = {J. Fluid Mech.},
	author = {Rouhi, A. and Lohse, D. and Marusic, I. and Sun, C. and Chung, D.},
	month = mar,
	year = {2021},
	pages = {A32}
}

@article{zhang_statistics_2017,
	title = {Statistics of kinetic and thermal energy dissipation rates in two-dimensional turbulent {Rayleigh}-{Bénard} convection},
	volume = {814},
	issn = {0022-1120, 1469-7645},
	url = {https://www.cambridge.org/core/product/identifier/S0022112017000192/type/journal_article},
	doi = {10.1017/jfm.2017.19},
	urldate = {2022-12-08},
	journal = {J. Fluid Mech.},
	author = {Zhang, Y. and Zhou, Q. and Sun, C.},
	month = mar,
	year = {2017},
	pages = {165--184},
}

@article{silano_numerical_2010,
	title = {Numerical simulations of {Rayleigh}-{Bénard} convection for {Prandtl} numbers between $10^{\textrm{-1}}$ and $10^{\textrm{4}}$ and {Rayleigh} numbers between $10^{\textrm{5}}$ and $10^{\textrm{9}}$},
	volume = {662},
	issn = {0022-1120, 1469-7645},
	url = {https://www.cambridge.org/core/product/identifier/S0022112010003290/type/journal_article},
	doi = {10.1017/S0022112010003290},
	urldate = {2022-12-08},
	journal = {J. Fluid Mech.},
	author = {Silano, G. and Sreenivasan, K. R. and Verzicco, R.},
	month = nov,
	year = {2010},
	pages = {409--446}
}

@article{yerragolam_how_2022,
	title = {How small-scale flow structures affect the heat transport in sheared thermal convection},
	volume = {944},
	issn = {0022-1120, 1469-7645},
	doi = {10.1017/jfm.2022.425},
	language = {en},
	urldate = {2023-07-01},
	journal = {J. Fluid Mech.},
	author = {Yerragolam, G. S. and Verzicco, R. and Lohse, D. and Stevens, R.J.A.M.},
	year = {2022},
	pages = {A1}
}

\end{document}